\documentclass[final,5p,times,authoryear]{elsarticle}

\usepackage{amsmath,amssymb,amsfonts,mathtools}
\usepackage{amsthm}
\usepackage{booktabs}
\usepackage{tikz}
\usepackage{pgfplots}
\usepgfplotslibrary{groupplots}
\usetikzlibrary{decorations.pathreplacing}
\pgfplotsset{compat=1.18}

\journal{Automatica}
\biboptions{authoryear,round}

\newtheorem{assumption}{Assumption}
\newtheorem{theorem}{Theorem}
\newtheorem{lemma}{Lemma}
\newtheorem{proposition}{Proposition}
\newtheorem{corollary}{Corollary}
\newtheorem{remark}{Remark}

\newcommand{\R}{\mathbb{R}}
\newcommand{\Sgn}{\operatorname{Sgn}}
\newcommand{\sgn}{\operatorname{sgn}}
\newcommand{\col}{\operatorname{col}}
\newcommand{\eps}{\varepsilon}
\newcommand{\logp}{\operatorname{log}^{+}}

\begin{document}
\begin{frontmatter}

\title{Output-Feedback Integral Sliding-Mode Extremum Seeking with Super-Twisting Differentiation}

\author[umd]{Hyuntae Kim\corref{cor1}}
\ead{htkim@cdsl.kr}
\cortext[cor1]{Corresponding author.}
\address[umd]{Department of Mechanical Engineering, University of Maryland, College Park, MD 20742, USA}

\begin{abstract}
This paper develops a dither-free output-feedback extremum-seeking controller for pre-stabilized nonlinear dynamic systems with unknown steady-state performance maps. The controller uses only the measured performance, the applied parameter, and internal states. Within a local state--parameter domain, integral sliding mode enforces equality between the applied parameter rate and a prescribed nonzero search velocity after finite-time capture under bounded matched rate-channel perturbations. This motion lets a super-twisting differentiator and rate normalization recover a gradient-sign surrogate without demodulation or averaging. A ramp-corrected moving-equilibrium coordinate cancels the first-order ramp-induced plant term and yields explicit flow, reversal, recovery, and dynamic-remainder bounds. The resulting local state--parameter tube supplies the regularity needed for finite-time exact differentiation. Combined with a deadbanded hybrid relay, these results establish local practical convergence to an explicit neighborhood of the unknown optimum. An 80-state tubular-reactor benchmark and paired held-out Monte Carlo tests compare a causal sampled realization with perturbation ESC and a rate-informed super-twisting comparator. The proposed realization succeeds in all 200 held-out cases and attains the smallest upper-tail objective gap under the stated tuning protocol. A matched-rate sweep connects performance degradation to applied-rate mismatch, near-stall motion, reversal, and regularization attenuation.
\end{abstract}


\end{frontmatter}

\section{Introduction}

Extremum-seeking control (ESC) adjusts an accessible parameter to optimize an unknown steady-state input--output characteristic. For nonlinear dynamic systems, a standard formulation assigns an equilibrium to each fixed parameter and optimizes the associated equilibrium performance map \citep{KrsticWang2000,AriyurKrstic2003,Scheinker2024}. Classical ESC commonly obtains gradient information through periodic perturbation, demodulation, and averaging or singular perturbation \citep{TanNesicMareels2006,Ghaffari2012,Duerr2013}. The perturbation frequency, filtering, plant bandwidth, and admissible update rate then enter the design jointly.

Several alternatives reduce or remove sinusoidal probing. Bounded-rate ESC constrains the update velocity \citep{ScheinkerKrstic2014}. Least-squares methods estimate local slopes from stored data \citep{Hunnekens2014}, while switching, self-perturbing, and sliding-mode methods use nonsinusoidal searches \citep{Angulo2015,Salsbury2020}. Integral sliding has been combined with synchronous detection for dynamic optimization and constrained costs \citep{Solis2019,Solis2021}; multivariable sliding-mode ESC instead uses cyclic searches and periodic switching \citep{AmindeOliveiraHsu2024}. Barrier-function integral sliding has also been used to reject matched perturbations around a modified sliding-mode seeker \citep{AlGhadiri2026}. Unit-vector ESC uses normalized discontinuous directions with sinusoidal probing and averaging \citep{Luo2026}. These approaches use stored-data slopes, periodic excitation, cyclic searches, or perturbation-rejection layers rather than the enforced-motion derivative ratio studied here.

Super-twisting differentiators provide finite-time estimates of signal derivatives \citep{Levant1998,Levant2003}. \citet{TorresZuniga2021} use super-twisting differentiation in extremum seeking, and dither-free derivative methods also exist for scalar static maps \citep{EfimovFridman2025}. For dynamic plants, however, the measured derivative combines the equilibrium gradient with a motion-induced transient. Exact differentiation also requires a bound on signal curvature. A derivative ratio adds another requirement: the applied parameter must actually move. Division by its rate fails at zero and becomes ill conditioned near zero. Regularization bounds the ratio but attenuates the gradient term as the rate vanishes. The analysis must therefore connect applied motion, plant transients, differentiator regularity, and the ratio interpretation.

The proposed dither-free output-feedback architecture combines integral sliding mode (ISM) with super-twisting (STW) differentiation and is denoted ISM--STW. The plant and any fixed low-level controller form an augmented state model, while the outer loop uses only the measured performance, the tuning parameter, and its own states. Integral sliding enforces a fixed-magnitude bidirectional search motion despite matched perturbations; a rate-mismatch proposition then quantifies the loss of gradient-sign resolution under stall or reversal. A ramp-corrected moving-equilibrium coordinate removes the leading constant-rate forcing, and Lyapunov flow and reversal bounds yield a uniform recovery estimate. The local state--parameter tube provides the curvature bound for exact super-twisting differentiation. The normalized ratio error separates a dynamic-lag contribution from a sensing-rate contribution, and a deadbanded relay gives an explicit local practical convergence radius. A linear gradient-growth condition is used only to sharpen the distance bound.

\citet{KimIFAC2026} considered an unknown static map with first-order actuation and a bounded-rate upstream command in the conference version. The present article places integral sliding in the applied parameter-rate channel and extends the setting to nonlinear dynamic systems with unknown equilibrium maps. It replaces assumed differentiation and settled-flow tubes by direct bounds for moving equilibria, reversals, recovery, and differentiator regularity. New results include the rate-mismatch proposition, the local practical convergence theorem, and the 80-state reactor comparison.

The analysis considers one tunable parameter, for which enforced nonzero motion permits recovery of the scalar gradient sign from an STW estimate of the derivative of one measured output after plant recovery. Validation uses a standard tubular-reactor benchmark. The model follows \citet{TrollbergJacobsen2018}; its operating-point-dependent phase lag and the multiple periodic stationary solutions reported for classical ESC make it a demanding benchmark. A common protocol with separate training, validation, and test sets compares the proposed controller with perturbation ESC (P-ESC), using fixed-gain and tuned versions denoted P-ESC-F and P-ESC-T, and with a rate-informed STW gradient method. The causal sampled realization of the proposed ISM--STW controller is denoted Causal ISM--STW. A matched-rate sweep varies the disturbance level through mismatch, near-stall, and reversal regimes and reports optimization success, upper-tail objective gap, applied-rate mismatch, and conditional regularization attenuation.

Throughout, \(\|\cdot\|\) denotes the Euclidean vector norm and its induced matrix norm, \(I\) is the identity matrix, and \(D^+\) is the upper Dini derivative. For a symmetric matrix, \(\lambda_{\min}\) and \(\lambda_{\max}\) denote its smallest and largest eigenvalues, and \(X\preceq Y\) denotes the Loewner order for symmetric matrices. Ordinary signs satisfy \(\sgn(0)=0\) and \(\sgn(a)=a/|a|\) for \(a\ne0\), while \(\Sgn(0)=[-1,1]\) and \(\Sgn(a)=\{\sgn(a)\}\) for \(a\ne0\). Primes denote derivatives with respect to the scalar parameter \(\theta\), subscripts on smooth maps, such as \(f_x\) and \(f_\theta\), denote partial derivatives, and superscripts \((-),(+)\) denote left and right limits at a hybrid event. For nonempty sets \(\mathcal A,\mathcal B\subset\R\), \(\operatorname{dist}(\mathcal A,\mathcal B):=\inf\{|a-b|:a\in\mathcal A,\ b\in\mathcal B\}\), with the singleton abbreviation \(\operatorname{dist}(a,\mathcal B)\). The notation \(\col(a,b)\) stacks its arguments.

\section{Dynamic extremum-seeking problem}
\label{sec:problem}

\subsection{Plant and equilibrium performance map}

Consider
\begin{equation}
  \dot x=f(x,\theta),
  \qquad
  y=h(x)+\nu,
  \label{eq:plant}
\end{equation}
where \(x\in\R^n\) is the state, \(\theta\in\R\) is a tunable parameter, \(y\in\R\) is the measured performance, and \(\nu\in\R\) is an additive sensing disturbance. The state may include both the physical plant and a fixed low-level output-feedback controller. The applied parameter \(\theta\) is available to the outer loop. The outer ESC uses \(y\), \(\theta\), and its own internal states, but no plant state.

Let \(\Theta=[\underline\theta,\overline\theta]\subset\R\) be a compact parameter interval. For every fixed \(\theta\in\Theta\), let \(\ell(\theta)\in\R^n\) denote an equilibrium satisfying
\begin{equation}
  f(\ell(\theta),\theta)=0.
  \label{eq:equilibrium_map}
\end{equation}
The unknown equilibrium performance map is
\begin{equation}
  J(\theta):=h(\ell(\theta)).
  \label{eq:J_def}
\end{equation}
The objective is to steer \(\theta\) to a neighborhood of the unknown minimizer \(\theta^\star\) of \(J\), using the measured performance and the available applied parameter. The maps \(\ell\), \(J\), and \(J'\) are analysis objects rather than online controller inputs.

\subsection{Assumptions}

\begin{assumption}[Smooth uniformly stable equilibria]
\label{ass:plant}
The function \(f\) is twice continuously differentiable on a neighborhood of the equilibrium graph
\[
  \mathcal M:=\{(\ell(\theta),\theta):\theta\in\Theta\},
\]
while \(h\) is twice continuously differentiable on a neighborhood of the state set \(\ell(\Theta)\), and \(\ell\) is twice continuously differentiable on a neighborhood of \(\Theta\). Define
\begin{equation}
  A(\theta):=f_x(\ell(\theta),\theta).
  \label{eq:A_def}
\end{equation}
There exist a continuously differentiable matrix function \(P(\theta)=P(\theta)^\top\) and constants \(\underline p,\overline p,q_P>0\) and \(\overline p_1\ge0\) such that, for all \(\theta\in\Theta\),
\begin{align}
  \underline p I&\preceq P(\theta)\preceq\overline p I,
  \label{eq:P_bounds}\\
  A(\theta)^\top P(\theta)+P(\theta)A(\theta)&\preceq-q_PI,
  \label{eq:lyap_frozen}\\
  \|P'(\theta)\|&\le\overline p_1.
  \label{eq:Pprime_bound}
\end{align}
The radius \(r_x>0\) is chosen so that the paired tube
\[
 \{(\ell(\theta)+e,\theta):\theta\in\Theta,\ \|e\|\le r_x\}
\]
lies in the smoothness neighborhood of \(f\), and its state projection lies in the smoothness neighborhood of \(h\). There also exists \(L_f>0\) such that, for every \(\theta\in\Theta\) and every \(e\in\R^n\) with \(\|e\|\le r_x\),
\begin{equation}
\begin{split}
 &\|f(\ell(\theta)+e,\theta)-A(\theta)e\|\\
 &\le\frac{L_f}{2}\|e\|^2
\end{split}
\label{eq:f_remainder}
\end{equation}
\end{assumption}

Assumption~\ref{ass:plant} makes \(A(\theta)\) nonsingular. Define
\begin{equation}
  \chi(\theta):=A(\theta)^{-1}\ell'(\theta),
  \label{eq:chi_def}
\end{equation}
and the finite constants
\begin{equation}
  \overline\chi:=\max_{\theta\in\Theta}\|\chi(\theta)\|,
  \qquad
  \overline\chi_1:=\max_{\theta\in\Theta}\|\chi'(\theta)\|.
  \label{eq:chi_bounds}
\end{equation}

\begin{assumption}[Detectable strict minimum]
\label{ass:map}
The map \(J\) has a unique minimizer \(\theta^\star\in\operatorname{int}\Theta\), satisfies \(J'(\theta^\star)=0\), and obeys
\begin{equation}
  \sgn J'(\theta)=\sgn(\theta-\theta^\star)
  \quad (\theta\ne\theta^\star).
  \label{eq:sign_detectability}
\end{equation}
Define
\begin{align}
  r_\Theta&:=\min\{\theta^\star-\underline\theta,
  \overline\theta-\theta^\star\},
  \label{eq:rTheta}\\
  R_\Theta&:=\max\{\theta^\star-\underline\theta,
  \overline\theta-\theta^\star\}.
  \label{eq:RTheta}
\end{align}
There is a continuous, strictly increasing function
\[
  \alpha_J:[0,R_\Theta]\to\R_{\ge0},
  \qquad
  \alpha_J(0)=0,
\]
such that
\begin{equation}
  |J'(\theta)|\ge\alpha_J(|\theta-\theta^\star|)
  \label{eq:alpha_J}
\end{equation}
for every \(\theta\in\Theta\). The derivative \(J'\) is \(L_J\)-Lipschitz on \(\Theta\).
\end{assumption}

\begin{assumption}[Disturbance regularity]
\label{ass:dist}
The additive sensing disturbance belongs to \(W^{2,\infty}\) and satisfies
\begin{equation}
  \|\dot\nu\|_{L_\infty}\le\overline\nu_1,
  \qquad
  \|\ddot\nu\|_{L_\infty}\le\overline\nu_2.
  \label{eq:nu_bounds}
\end{equation}
The matched perturbation \(d_\theta\) in the parameter-rate channel satisfies
\begin{equation}
  |d_\theta(t)|\le\overline d_\theta
  \quad\text{for almost all }t\ge0.
  \label{eq:dtheta_bound}
\end{equation}
\end{assumption}

\section{Output-feedback ISM--STW extremum seeker}
\label{sec:controller}

\subsection{Integral-sliding search motion}

The tunable parameter is implemented through
\begin{equation}
  \dot\theta=u_\theta+d_\theta.
  \label{eq:parameter_channel}
\end{equation}
Here \(u_\theta\) is the outer-loop rate command. This channel represents an embedded command interface. The applied value \(\theta\) is available to the controller, whereas \(d_\theta\) is not measured. The nominal software-update case is recovered by setting \(d_\theta\equiv0\).
The commanded search velocity satisfies
\begin{equation}
  v(t)\in\{-\rho,+\rho\},
  \qquad \rho>0,
  \label{eq:v_values}
\end{equation}
and is constant between relay reversals. Introduce
\begin{equation}
  \dot\xi=v,
  \qquad
  s:=\theta-\xi,
  \label{eq:xi_s}
\end{equation}
and apply
\begin{equation}
  u_\theta\in v-k_I\Sgn(s),
  \qquad k_I>0.
  \label{eq:ism_update}
\end{equation}
The resulting differential inclusion is interpreted in the Filippov sense \citep{Filippov1988}. The controller state is initialized by
\begin{equation}
  \xi(0)=\theta(0).
  \label{eq:integral_initialization}
\end{equation}
This initialization leaves the plant state \(x(0)\) arbitrary and gives \(s(0)=0\). Equivalently,
\[
 s(t)=\theta(t)-\theta(0)-\int_0^t v(\tau)\,d\tau,
\]
so the integral manifold contains the initial condition and has no reaching phase under the standard initialization.

\begin{proposition}[Exact fixed-rate motion]
\label{prop:ism}
If \(k_I>\overline d_\theta\), the reaching time \(T_I\) from an arbitrary initial sliding error satisfies
\begin{equation}
  T_I\le\frac{|s(0)|}{k_I-\overline d_\theta}.
  \label{eq:TI}
\end{equation}
After reaching,
\begin{equation}
  s(t)=0,
  \qquad
  \dot\theta(t)=v(t)
  \label{eq:theta_dot_v}
\end{equation}
for almost all subsequent times. Under \eqref{eq:integral_initialization}, \eqref{eq:theta_dot_v} holds from \(t=0\).
\end{proposition}

\begin{proof}
Equations \eqref{eq:parameter_channel}--\eqref{eq:ism_update} give
\[
  \dot s+k_I\Sgn(s)\ni d_\theta.
\]
For \(s\ne0\),
\[
  D^+|s|\le-(k_I-\overline d_\theta),
\]
which proves \eqref{eq:TI}. The operator \(k_I\Sgn\) is maximal monotone, so the inclusion has a unique absolutely continuous solution for each admissible \(d_\theta\). At \(s=0\), the selection \(d_\theta/k_I\in[-1,1]\) makes \(\dot s=0\); uniqueness therefore makes \(s=0\) invariant. Equation \eqref{eq:theta_dot_v} follows from \(s=\theta-\xi\).
\end{proof}

\subsection{Super-twisting output differentiator}

Let
\begin{equation}
  e_y:=\widehat y-y.
  \label{eq:ey_def}
\end{equation}
The differentiator is
\begin{align}
  \dot{\widehat y}
  &\in-\lambda_1|e_y|^{1/2}\Sgn(e_y)+w,
  \label{eq:stw_yhat}\\
  \dot w
  &\in-\lambda_2\Sgn(e_y),
  \label{eq:stw_w}
\end{align}
where \(\lambda_1,\lambda_2>0\). The state \(w\) is the derivative estimate supplied to the ratio surrogate.

\subsection{Derivative ratio and hybrid relay}

The flow-time gradient surrogate is
\begin{equation}
  \widehat g:=\frac{w}{v}.
  \label{eq:gradient_surrogate}
\end{equation}
The denominator is known and nonzero because \(|v|=\rho\). A gradient interpretation also requires the applied parameter to follow the prescribed direction. Proposition~\ref{prop:ism} gives \(\dot\theta=v\), so after sliding capture the prescribed denominator equals the applied rate. Let \(h_g>0\) be the deadband, \(\tau_{\min}>0\) the minimum dwell, and \(T_A\ge0\) the activation time. The dwell clock \(\tau_d\) evolves as
\begin{equation}
  \dot\tau_d=1,
  \qquad
  \tau_d(0)=\tau_{\min}.
  \label{eq:dwell_clock}
\end{equation}
Before \(T_A\), the initial velocity \(v_0\in\{-\rho,+\rho\}\) is held. For \(t\ge T_A\), a relay reversal is permitted when
\begin{equation}
  \tau_d\ge\tau_{\min},
  \qquad
  |\widehat g^-|>h_g,
  \qquad
  v^-\widehat g^->0.
  \label{eq:jump_set}
\end{equation}
At a reversal,
\begin{equation}
  v^+=-\rho\sgn(\widehat g^-),
  \qquad
  \tau_d^+=0.
  \label{eq:relay_update}
\end{equation}
At a relay reversal, the tuning parameter remains continuous while its prescribed rate changes sign.
The deadband makes \(\widehat g^-\ne0\), so the ordinary sign in \eqref{eq:relay_update} is well defined. Since \(v\widehat g=w\), the last condition in \eqref{eq:jump_set} identifies a positive estimated performance derivative along the current motion. Let \(t_k\), \(k\ge1\), denote the relay-reversal times generated by \eqref{eq:jump_set}--\eqref{eq:relay_update}; then \(t_{k+1}-t_k\ge\tau_{\min}\).

The outer-loop implementation uses \(y\), the available applied parameter \(\theta\), \((\widehat y,w)\), \((\xi,s)\), \(v\), and \(\tau_d\). The plant state and equilibrium map enter only the analysis; neither quantity is required online.

\section{Dynamic reduction under fixed-rate search}
\label{sec:dynamic_reduction}

This section derives the dynamic term in the derivative ratio directly from \eqref{eq:plant}.

\subsection{Ramp-corrected moving-equilibrium coordinate}

On every interval after integral sliding-mode capture and between relay reversals, Proposition~\ref{prop:ism} gives \(\dot\theta=v\) with constant \(v\in\{-\rho,+\rho\}\). Define
\begin{equation}
  \eta:=x-\ell(\theta)-v\chi(\theta),
  \label{eq:eta_def}
\end{equation}
where \(\chi\) is given by \eqref{eq:chi_def}. The term \(v\chi(\theta)\) is the first-order state lag generated by the constant parameter rate.

Using \(A(\theta)\chi(\theta)=\ell'(\theta)\) gives
\begin{equation}
  \dot\eta=A(\theta)\eta+R_f(\eta,v,\theta)-v^2\chi'(\theta),
  \label{eq:eta_dynamics}
\end{equation}
where
\begin{equation}
\begin{split}
 R_f(\eta,v,\theta):={}&f\bigl(\ell(\theta)+v\chi(\theta)+\eta,\theta\bigr)\\
 &-A(\theta)\bigl(v\chi(\theta)+\eta\bigr).
\end{split}
\label{eq:Rf_def}
\end{equation}
Whenever \(\|\eta\|+\rho\overline\chi\le r_x\), Assumption~\ref{ass:plant} gives
\begin{equation}
  \|R_f\|\le\frac{L_f}{2}
  (\|\eta\|+\rho\overline\chi)^2.
  \label{eq:Rf_bound}
\end{equation}

\subsection{Flow estimate}

Choose \(r_\eta>0\) and \(\rho_0>0\) such that
\begin{equation}
  r_\eta+\rho_0\overline\chi\le r_x.
  \label{eq:local_radius}
\end{equation}
Define
\begin{align}
 c_0:={}&q_P-\rho_0\overline p_1-\overline pL_fr_\eta
 -2\overline pL_f\rho_0\overline\chi,
 \label{eq:c0_def}\\
 c_f:={}&\overline pL_f\overline\chi^2
 +2\overline p\overline\chi_1.
 \label{eq:cf_def}
\end{align}
The local design region is selected so that
\begin{equation}
  c_0>0.
  \label{eq:c0_positive}
\end{equation}
Set
\begin{equation}
  \lambda_\eta:=\frac{c_0}{2\overline p},
  \qquad
  b_\eta:=\frac{c_f^2}{2c_0},
  \qquad
  \beta_\eta:=\frac{b_\eta}{\lambda_\eta}.
  \label{eq:eta_constants}
\end{equation}

\begin{lemma}[Ramp-flow contraction]
\label{lem:flow}
Let
\begin{equation}
  V_\eta:=\eta^\top P(\theta)\eta.
  \label{eq:Veta_def}
\end{equation}
If \(0<\rho\le\rho_0\) and \(\|\eta\|\le r_\eta\), then
\begin{equation}
  \dot V_\eta\le-\lambda_\eta V_\eta+b_\eta\rho^4.
  \label{eq:V_flow}
\end{equation}
Consequently, for any flow interval beginning at \(t_a\),
\begin{equation}
\begin{split}
 V_\eta(t)\le{}&e^{-\lambda_\eta(t-t_a)}V_\eta(t_a)\\
 &+\beta_\eta\rho^4\bigl(1-e^{-\lambda_\eta(t-t_a)}\bigr)\\
 \le{}&\max\{V_\eta(t_a),\beta_\eta\rho^4\}.
\end{split}
\label{eq:V_flow_solution}
\end{equation}
\end{lemma}

\begin{proof}
Differentiating \eqref{eq:Veta_def} along \eqref{eq:eta_dynamics} and using \(\dot\theta=v\) gives
\begin{equation}
\begin{split}
 \dot V_\eta={}&\eta^\top(A^\top P+PA)\eta
 +v\eta^\top P'\eta\\
 &+2\eta^\top P(R_f-v^2\chi').
\end{split}
\label{eq:Veta_expanded}
\end{equation}
Equations \eqref{eq:lyap_frozen}, \eqref{eq:Pprime_bound}, and \eqref{eq:Rf_bound} imply
\begin{equation}
  \dot V_\eta\le-c_0\|\eta\|^2+c_f\rho^2\|\eta\|.
  \label{eq:Veta_bound}
\end{equation}
Young's inequality yields
\[
  c_f\rho^2\|\eta\|
  \le\frac{c_0}{2}\|\eta\|^2+\frac{c_f^2}{2c_0}\rho^4.
\]
Since \(V_\eta\le\overline p\|\eta\|^2\), one has \(-\tfrac{c_0}{2}\|\eta\|^2\le-\tfrac{c_0}{2\overline p}V_\eta=-\lambda_\eta V_\eta\). Together with the definition of \(b_\eta\), this proves \eqref{eq:V_flow}. The scalar comparison equation then gives \eqref{eq:V_flow_solution}.
\end{proof}

\subsection{Reversal jump and recovery}

At a relay reversal time \(t_k\), \(x\) and \(\theta\) are continuous and \(v^+=-v^-\). Hence
\begin{equation}
  \eta(t_k^+)=\eta(t_k^-)+2v^-\chi(\theta(t_k)).
  \label{eq:eta_jump_exact}
\end{equation}
For any \(\eps_s>0\), define
\begin{equation}
  \mu_s:=1+\eps_s,
  \qquad
  \kappa_s:=4\left(1+\frac{1}{\eps_s}\right)
  \overline p\,\overline\chi^2.
  \label{eq:jump_constants}
\end{equation}
Then
\begin{equation}
  V_\eta(t_k^+)\le\mu_sV_\eta(t_k^-)+\kappa_s\rho^2.
  \label{eq:V_jump}
\end{equation}

Define
\begin{equation}
  q_s:=\mu_s e^{-\lambda_\eta\tau_{\min}}.
  \label{eq:qs_def}
\end{equation}
The switch-to-switch contraction condition is
\begin{equation}
  q_s<1.
  \label{eq:qs_condition}
\end{equation}
Let \(V_{\eta,k}^+:=V_\eta(t_k^+)\). Combining \eqref{eq:V_flow_solution} and \eqref{eq:V_jump}, and using \(1-e^{-\lambda_\eta(t_{k+1}-t_k)}\le1\), gives, for every pair of consecutive reversals,
\begin{equation}
  V_{\eta,k+1}^+
  \le q_sV_{\eta,k}^+
  +\mu_s\beta_\eta\rho^4+\kappa_s\rho^2.
  \label{eq:V_recursion}
\end{equation}

Choose any \(C_V>\beta_\eta\). An offline-selected activation time \(T_A\) is chosen so that the initial velocity is held long enough to ensure
\begin{equation}
  V_\eta(T_A^-)\le C_V\rho^4.
  \label{eq:startup_recovered}
\end{equation}
At the first reversal \(t_1\ge T_A\), \eqref{eq:V_flow_solution}, \eqref{eq:startup_recovered}, and \(C_V>\beta_\eta\) give
\[
  V_\eta(t_1^-)\le C_V\rho^4.
\]
Hence
\begin{equation}
  V_{\eta,1}^+
  \le \mu_sC_V\rho^4+\kappa_s\rho^2.
  \label{eq:first_jump_bound}
\end{equation}
Accordingly, define
\begin{equation}
\begin{split}
 \overline V_\eta^+:=\max\biggl\{&\mu_sC_V\rho^4+\kappa_s\rho^2,\\
 &\frac{\mu_s\beta_\eta\rho^4+\kappa_s\rho^2}{1-q_s}\biggr\}.
\end{split}
\label{eq:Vbar}
\end{equation}
Then \(V_{\eta,k}^+\le\overline V_\eta^+\) for every reversal. The local flow estimate is self-consistent when
\begin{equation}
  \max\{\overline V_\eta^+,\beta_\eta\rho^4\}
  \le\underline p r_\eta^2.
  \label{eq:local_consistency}
\end{equation}

For \(a\ge0\), define \(\logp(a):=\log(\max\{1,a\})\). A recovery time satisfying
\begin{equation}
  \tau_r\ge\frac{1}{\lambda_\eta}
  \logp\!\left(
  \frac{\overline V_\eta^+}{(C_V-\beta_\eta)\rho^4}
  \right)
  \label{eq:tau_r}
\end{equation}
gives
\begin{equation}
  V_\eta(t)\le C_V\rho^4,
  \qquad
  \|\eta(t)\|\le C_\eta\rho^2
  \label{eq:eta_recovered}
\end{equation}
for \(t\in[t_k+\tau_r,t_{k+1})\), where
\begin{equation}
  C_\eta:=\sqrt{\frac{C_V}{\underline p}}.
  \label{eq:Ceta}
\end{equation}
The dwell is chosen to satisfy
\begin{equation}
  \tau_{\min}\ge\tau_r.
  \label{eq:dwell_recovery}
\end{equation}

Fix an admissible set of initial triples \((x(0),\theta(0),v_0)\). Assume that the known constant \(\overline V_0\) satisfies \(V_\eta(0)\le\overline V_0\) throughout this set. Under \eqref{eq:integral_initialization}, choosing the offline hold time
\begin{equation}
  \tau_0\ge\frac{1}{\lambda_\eta}
  \logp\!\left(
  \frac{\overline V_0}{(C_V-\beta_\eta)\rho^4}
  \right)
  \label{eq:tau0}
\end{equation}
ensures \eqref{eq:startup_recovered} for every admissible initial condition, without online access to the plant state. Proposition~\ref{prop:ism} adds the finite capture time \(T_I\) when the integral state is not initialized by \eqref{eq:integral_initialization}.

\subsection{Measured-derivative decomposition}

Define the paired state--parameter tube and its state projection by
\begin{align}
 \mathcal Z&:=\{(x,\theta)\in\R^n\times\Theta:
 \|x-\ell(\theta)\|\le r_x\},\label{eq:Z_def}\\
 \mathcal X&:=\{x\in\R^n:(x,\theta)\in\mathcal Z
 \text{ for some }\theta\in\Theta\}.\label{eq:X_def}
\end{align}
The following constants are finite by Assumption~\ref{ass:plant}:
\begin{equation}
\begin{gathered}
 \overline\ell_1:=\max_{\theta\in\Theta}\|\ell'(\theta)\|,
 \qquad
 \overline A:=\max_{\theta\in\Theta}\|A(\theta)\|,\\
 \overline h_1:=\max_{x\in\mathcal X}\|h_x(x)\|,
 \qquad
 L_h:=\text{a Lipschitz constant of }h_x\text{ on }\mathcal X.
\end{gathered}
\label{eq:derivative_bounds_1}
\end{equation}
Since
\begin{equation}
  J'(\theta)=h_x(\ell(\theta))\ell'(\theta),
  \label{eq:Jprime}
\end{equation}
substitution of
\[
  f(x,\theta)=\ell'(\theta)v+A(\theta)\eta+R_f
\]
into \(\dot y=h_x(x)f(x,\theta)+\dot\nu\) gives
\begin{equation}
  \dot y=J'(\theta)v+\Delta_{\mathrm{dyn}}+\dot\nu,
  \label{eq:ydot_decomposition}
\end{equation}
where
\begin{equation}
\begin{split}
 \Delta_{\mathrm{dyn}}={}&v\bigl[h_x(x)-h_x(\ell(\theta))\bigr]\ell'(\theta)\\
 &+h_x(x)A(\theta)\eta+h_x(x)R_f.
\end{split}
\label{eq:Delta_def}
\end{equation}
Set
\begin{equation}
\begin{split}
 C_\Delta:={}&L_h\overline\ell_1
 (\overline\chi+C_\eta\rho_0)
 +\overline h_1\overline A C_\eta\\
 &+\frac{\overline h_1L_f}{2}
 (\overline\chi+C_\eta\rho_0)^2.
\end{split}
\label{eq:Cdelta}
\end{equation}

\begin{lemma}[Dynamic derivative remainder]
\label{lem:dynamic_remainder}
On every recovered portion \([t_k+\tau_r,t_{k+1})\),
\begin{equation}
  |\Delta_{\mathrm{dyn}}(t)|\le C_\Delta\rho^2.
  \label{eq:Delta_bound}
\end{equation}
\end{lemma}

\begin{proof}
Equations \eqref{eq:eta_def} and \eqref{eq:eta_recovered} give
\[
  \|x-\ell(\theta)\|
  \le\rho\overline\chi+C_\eta\rho^2.
\]
The first term in \eqref{eq:Delta_def} is therefore bounded by
\[
 \rho L_h(\rho\overline\chi+C_\eta\rho^2)\overline\ell_1.
\]
The second term is at most \(\overline h_1\overline A C_\eta\rho^2\). Moreover, \eqref{eq:Rf_bound} and \(\|\eta\|\le C_\eta\rho^2\) give
\[
 \|R_f\|\le\frac{L_f}{2}\rho^2
 (\overline\chi+C_\eta\rho)^2,
\]
so the third term is at most \(\tfrac{\overline h_1L_f}{2}\rho^2(\overline\chi+C_\eta\rho)^2\). Factoring \(\rho^2\), using \(\rho\le\rho_0\), and collecting the three coefficients gives \eqref{eq:Delta_bound} with \eqref{eq:Cdelta}.
\end{proof}

\section{Endogenous exact differentiation}
\label{sec:stw_analysis}

The tube \(\mathcal Z\) also provides the regularity required by the super-twisting differentiator. Define
\begin{equation}
\begin{gathered}
 \overline f:=\max_{(x,\theta)\in\mathcal Z}\|f(x,\theta)\|,
 \quad
 \overline f_x:=\max_{(x,\theta)\in\mathcal Z}\|f_x(x,\theta)\|,\\
 \overline f_\theta:=\max_{(x,\theta)\in\mathcal Z}\|f_\theta(x,\theta)\|,
 \quad
 \overline h_2:=\max_{x\in\mathcal X}\|h_{xx}(x)\|.
\end{gathered}
\label{eq:compact_bounds}
\end{equation}
Define
\begin{equation}
  \overline\theta_1:=\rho+k_I+\overline d_\theta.
  \label{eq:theta_rate_bound}
\end{equation}
Before and after integral sliding-mode capture,
\begin{equation}
  |\dot\theta|\le\overline\theta_1.
  \label{eq:theta_rate_global}
\end{equation}
For almost all flow times, the chain rule gives
\begin{equation}
 \ddot y=f^\top h_{xx}f
 +h_x\bigl(f_xf+f_\theta\dot\theta\bigr)+\ddot\nu.
 \label{eq:yddot_chain}
\end{equation}
Consequently,
\begin{equation}
\begin{split}
 |\ddot y|\le{}&\overline h_2\overline f^2
 +\overline h_1
 (\overline f_x\overline f+\overline f_\theta\overline\theta_1)
 +\overline\nu_2\\
 =:{}&Y_2.
\end{split}
\label{eq:Y2}
\end{equation}
At a relay reversal, \(x\) and \(\theta\) remain continuous. Assumption~\ref{ass:dist} provides a Lipschitz representative of \(\dot\nu\). Hence \(\dot y\) is continuous and the essential bound \eqref{eq:Y2} remains valid.

For \(\delta\in[-Y_2,Y_2]\), define
\begin{equation}
 M_D(\delta):=
 \begin{bmatrix}
 -\lambda_1/2 & 1/2\\
 -(\lambda_2+\delta) & 0
 \end{bmatrix}.
 \label{eq:MD}
\end{equation}
Choose \(P_D=P_D^\top\succ0\) and \(Q_D=Q_D^\top\succ0\) such that
\begin{equation}
 M_D(\delta_i)^\top P_D+P_DM_D(\delta_i)
 \preceq-Q_D,
 \quad
 \delta_i\in\{-Y_2,+Y_2\}.
 \label{eq:stw_lmi}
\end{equation}
Since the left-hand side is affine in \(\delta\), the same inequality holds for every \(\delta\in[-Y_2,Y_2]\).

\begin{lemma}[Finite-time exact output differentiation]
\label{lem:stw}
Along a trajectory that remains in \(\mathcal Z\) and satisfies the rate bound \eqref{eq:theta_rate_global}, suppose \eqref{eq:stw_lmi} holds. Define
\begin{equation}
  \widetilde w:=w-\dot y,
  \qquad
  \zeta_1:=|e_y|^{1/2}\sgn(e_y),
  \qquad
  \zeta_D:=\col(\zeta_1,\widetilde w),
  \label{eq:stw_coordinates}
\end{equation}
and \(V_D:=\zeta_D^\top P_D\zeta_D\). Then
\begin{equation}
  \dot V_D\le-c_D\sqrt{V_D},
  \qquad
  c_D:=\frac{\lambda_{\min}(Q_D)}
  {\sqrt{\lambda_{\max}(P_D)}}.
  \label{eq:VD_decay}
\end{equation}
Consequently, any certified differentiation time \(T_D\) satisfying
\begin{equation}
  T_D\ge\frac{2\sqrt{V_D(0)}}{c_D}
  \label{eq:TD_general}
\end{equation}
ensures
\begin{equation}
  \widehat y(t)=y(t),
  \qquad
  w(t)=\dot y(t)
  \label{eq:exact_diff}
\end{equation}
for all \(t\ge T_D\). Once reached, \eqref{eq:exact_diff} is preserved across relay reversals.
\end{lemma}

\begin{proof}
For \(e_y\ne0\), the coordinates in \eqref{eq:stw_coordinates} satisfy
\begin{equation}
  \dot\zeta_D=\frac{1}{|\zeta_1|}M_D(\delta)\zeta_D,
  \qquad
  \delta:=\ddot y\,\sgn(e_y),
  \label{eq:z_dynamics}
\end{equation}
with \(|\delta|\le Y_2\). Hence
\begin{equation}
\begin{split}
 \dot V_D
 &=\frac{1}{|\zeta_1|}\zeta_D^\top
 (M_D^\top P_D+P_DM_D)\zeta_D\\
 &\le-\frac{\lambda_{\min}(Q_D)}{|\zeta_1|}\|\zeta_D\|^2.
\end{split}
\label{eq:VD_step}
\end{equation}
Since \(|\zeta_1|\le\|\zeta_D\|\) and
\(V_D\le\lambda_{\max}(P_D)\|\zeta_D\|^2\), inequality \eqref{eq:VD_decay} follows. Integrating it shows that the exact manifold is reached no later than \(2\sqrt{V_D(0)}/c_D\); hence any \(T_D\) satisfying \eqref{eq:TD_general} is sufficient. The Filippov inclusion extends the result to \(e_y=0\) and makes the exact manifold invariant \citep{Levant1998,MorenoOsorio2012}. Continuity of \(\dot y\) at a relay reversal preserves that manifold.
\end{proof}

With \(\widehat y(0)=y(0)\) and \(w(0)=0\), define
\begin{equation}
  Y_1:=\overline h_1\overline f+\overline\nu_1.
  \label{eq:Y1}
\end{equation}
Then it is sufficient to choose
\begin{equation}
  T_D\ge
  \frac{2\lambda_{\max}(P_D)Y_1}
  {\lambda_{\min}(Q_D)}.
  \label{eq:TD_explicit}
\end{equation}
Under the standard initialization \eqref{eq:integral_initialization}, select the relay-activation time as
\begin{equation}
  T_A\ge\max\{\tau_0,T_D\}.
  \label{eq:TA}
\end{equation}
For a nonstandard integral-state initialization, the startup estimate is restarted at the capture time \(T_I\) with \(V_\eta(T_I)\), and \(T_I\) is added to the resulting hold time.

\section{Gradient-sign recovery and practical convergence}
\label{sec:convergence}

The effect of imperfect rate realization can be separated from the plant and differentiator terms before specializing to the proposed closed loop.

\begin{proposition}[Applied-rate mismatch and gradient-sign resolution]
\label{prop:rate_mismatch}
On a flow interval with \(|v|=\rho\), define
\begin{equation}
 e_v:=\dot\theta-v,
 \qquad
 r_y:=w-J'(\theta)\dot\theta.
 \label{eq:ev_ry}
\end{equation}
Then
\begin{equation}
 \widehat g
 =J'(\theta)\left(1+\frac{e_v}{v}\right)+\frac{r_y}{v}.
 \label{eq:rate_mismatch_identity}
\end{equation}
Suppose that \(|e_v|\le\overline e_v\) and \(|r_y|\le\overline r_y\). If \(\overline e_v<\rho\), then
\begin{equation}
 |J'(\theta)|>
 \frac{\overline r_y+\rho h_g}{\rho-\overline e_v}
 \label{eq:mismatch_sign_threshold}
\end{equation}
implies \(|\widehat g|>h_g\) and
\(\sgn\widehat g=\sgn J'(\theta)\). If \(\overline e_v\ge\rho\), these bounds alone cannot provide a uniform gradient-sign guarantee.
\end{proposition}

\begin{proof}
Multiplying \eqref{eq:rate_mismatch_identity} by \(\sgn J'(\theta)\) gives
\[
 \sgn J'(\theta)\widehat g
 \ge |J'(\theta)|\left(1-\frac{\overline e_v}{\rho}\right)
 -\frac{\overline r_y}{\rho}.
\]
Condition \eqref{eq:mismatch_sign_threshold} makes the right-hand side larger than \(h_g\), proving both conclusions. If \(\overline e_v\ge\rho\), the admissible choice \(e_v=-v\) gives \(\dot\theta=0\). With \(r_y=0\), \eqref{eq:rate_mismatch_identity} then gives \(\widehat g=0\) for an arbitrary nonzero \(J'(\theta)\), so a uniform sign guarantee is impossible.
\end{proof}

After \(T_A\), every admissible relay reversal occurs after the current plant transient has recovered because \(\tau_{\min}\ge\tau_r\). Proposition~\ref{prop:ism} gives \(e_v=0\), while Lemmas~\ref{lem:dynamic_remainder} and~\ref{lem:stw} give
\begin{equation}
  r_y=\Delta_{\mathrm{dyn}}+\dot\nu.
  \label{eq:ratio_residual_recovered}
\end{equation}
Consequently,
\begin{equation}
  |\widehat g-J'(\theta)|
  \le C_\Delta\rho+\frac{\overline\nu_1}{\rho}
  =:\eps_g(\rho).
  \label{eq:epsg}
\end{equation}
The first term is the normalized dynamic lag, and the second is the normalized sensing-rate disturbance.

\begin{corollary}[ISM-recovered gradient sign]
\label{cor:ism_sign}
On a recovered flow portion after \(T_A\),
\begin{equation}
  |J'(\theta)|>\eps_g(\rho)+h_g
  \label{eq:outside_gradient_residual}
\end{equation}
implies
\begin{equation}
  |\widehat g|>h_g,
  \qquad
  \sgn\widehat g=\sgn J'(\theta).
  \label{eq:sign_correct}
\end{equation}
\end{corollary}

\begin{proof}
Use Proposition~\ref{prop:rate_mismatch} with \(\overline e_v=0\) and \(\overline r_y=C_\Delta\rho^2+\overline\nu_1\).
\end{proof}

Define
\begin{equation}
  \eps_{\mathrm{tot}}(\rho):=\eps_g(\rho)+h_g.
  \label{eq:eps_total}
\end{equation}
Assume
\begin{equation}
  \eps_{\mathrm{tot}}(\rho)<\alpha_J(r_\Theta),
  \label{eq:residual_inside}
\end{equation}
and set
\begin{equation}
  R_\rho:=\alpha_J^{-1}(\eps_{\mathrm{tot}}(\rho)).
  \label{eq:Rrho}
\end{equation}
Then \(0<R_\rho<r_\Theta\). Define the residual interval
\begin{equation}
 \mathcal R_\rho:=[\theta^\star-R_\rho,\theta^\star+R_\rho].
 \label{eq:Rset}
\end{equation}
It lies strictly inside \(\Theta\), and Corollary~\ref{cor:ism_sign} applies whenever \(\theta\notin\mathcal R_\rho\).

If the velocity is aligned for minimization, then
\begin{equation}
  v=-\rho\sgn J'(\theta),
  \label{eq:aligned_velocity}
\end{equation}
and Proposition~\ref{prop:ism} gives
\begin{equation}
  D^+|\theta-\theta^\star|=-\rho.
  \label{eq:distance_descent}
\end{equation}
If the velocity is misaligned, the last two inequalities in \eqref{eq:jump_set} hold after recovery. The relay then restores \eqref{eq:aligned_velocity} as soon as the dwell clock permits a reversal. The remaining waiting time is at most \(\tau_{\min}\), so the largest outward parameter excursion is \(\rho\tau_{\min}\).

For a precise local-domain condition, let \(\Theta_0\) be the projection of the prescribed initial set onto its parameter coordinate and define
\begin{equation}
  \Delta_\Theta:=R_\rho+\rho(T_A+\tau_{\min})
  \label{eq:DeltaTheta}
\end{equation}
and, for each \(\theta_0\in\Theta_0\),
\begin{equation}
\begin{split}
 \mathcal I_\Theta(\theta_0):=\bigl[&\min\{\theta_0,\theta^\star\}-\Delta_\Theta,\\
 &\max\{\theta_0,\theta^\star\}+\Delta_\Theta\bigr].
\end{split}
\label{eq:ITheta}
\end{equation}
The uniform condition \(\mathcal I_\Theta(\theta_0)\subset\operatorname{int}\Theta\) for all \(\theta_0\in\Theta_0\) keeps every admissible startup and relay trajectory inside the parameter interval on which the constants above are defined.

\begin{theorem}[Local practical convergence]
\label{thm:main}
Suppose the following conditions hold.
\begin{enumerate}
\item Assumptions~\ref{ass:plant}--\ref{ass:dist} and the initialization \eqref{eq:integral_initialization} hold.
\item The local constants satisfy \eqref{eq:local_radius} and \eqref{eq:c0_positive} for some \(0<\rho\le\rho_0\).
\item The flow--jump design satisfies \eqref{eq:qs_condition}, \eqref{eq:local_consistency}, \eqref{eq:tau_r}, and \eqref{eq:dwell_recovery} for selected \(\eps_s>0\) and \(C_V>\beta_\eta\).
\item The gains satisfy \(k_I>\overline d_\theta\), the differentiator condition \eqref{eq:stw_lmi}, and the activation condition \eqref{eq:TA}.
\item The residual condition \eqref{eq:residual_inside} holds, and \(\mathcal I_\Theta(\theta_0)\subset\operatorname{int}\Theta\) for every \(\theta_0\in\Theta_0\).
\end{enumerate}
Assume additionally that
\begin{equation}
  \max\{\overline V_0,\beta_\eta\rho^4\}
  \le\underline p r_\eta^2.
  \label{eq:initial_consistency}
\end{equation}
Then every initial condition in the prescribed initial set generates a forward-complete hybrid Filippov trajectory. The pair \((x(t),\theta(t))\) remains in \(\mathcal Z\). Its relay reversals cannot accumulate, and
\begin{equation}
  \limsup_{t\to\infty}|\theta(t)-\theta^\star|
  \le R_\rho+\rho\tau_{\min}.
  \label{eq:main_parameter_bound}
\end{equation}
Moreover,
\begin{equation}
  \limsup_{t\to\infty}|J'(\theta(t))|
  \le L_J\bigl(R_\rho+\rho\tau_{\min}\bigr).
  \label{eq:main_gradient_bound}
\end{equation}
\end{theorem}

\begin{proof}
The dwell clock gives at least \(\tau_{\min}\) between consecutive reversals, so no finite accumulation is possible. Lemma~\ref{lem:flow}, \eqref{eq:initial_consistency}, the recursion \eqref{eq:V_recursion}, and \eqref{eq:local_consistency} keep the plant state in the local tube. For the selected initial condition, the inclusion \(\mathcal I_\Theta(\theta(0))\subset\operatorname{int}\Theta\) keeps \(\theta\) in \(\Theta\). Smoothness on these compact sets gives forward completeness. Lemma~\ref{lem:stw} establishes exact differentiation before relay activation and preserves it afterward.

After \(T_A\), Corollary~\ref{cor:ism_sign} gives the correct gradient sign whenever \(\theta\notin\mathcal R_\rho\). If the held velocity is aligned, \eqref{eq:distance_descent} decreases \(\operatorname{dist}(\theta,\mathcal R_\rho)\) at rate \(\rho\). If it is misaligned, \eqref{eq:sign_detectability} prevents the true gradient sign from changing before the trajectory re-enters \(\mathcal R_\rho\). The jump condition therefore persists after recovery, and the dwell clock permits a corrective reversal within at most \(\tau_{\min}\). During that waiting interval, \(|\dot\theta|=\rho\), so the outward excursion is at most \(\rho\tau_{\min}\). Every excursion outside \(\mathcal R_\rho\) is thus followed by an aligned segment that decreases the distance, and no excursion can exceed \(R_\rho+\rho\tau_{\min}\). This proves \eqref{eq:main_parameter_bound}.

Since \(J'(\theta^\star)=0\) and \(J'\) is \(L_J\)-Lipschitz, \(|J'(\theta)|\le L_J|\theta-\theta^\star|\). Combining this inequality with \eqref{eq:main_parameter_bound} proves \eqref{eq:main_gradient_bound}.
\end{proof}

\begin{corollary}[Linear gradient-growth condition]
\label{cor:strong}
If, in addition, there exists \(m>0\) such that
\begin{equation}
  |J'(\theta)|\ge m|\theta-\theta^\star|,
  \qquad \text{for all }\theta\in\Theta,
  \label{eq:gradient_growth}
\end{equation}
then
\begin{equation}
\begin{split}
 \limsup_{t\to\infty}|\theta(t)-\theta^\star|
 \le{}&\frac{1}{m}
 \left(C_\Delta\rho+\frac{\overline\nu_1}{\rho}+h_g\right)\\
 &+\rho\tau_{\min}.
\end{split}
\label{eq:strong_parameter_bound}
\end{equation}
\end{corollary}

\begin{proof}
Under \eqref{eq:gradient_growth}, the gradient-growth function in Assumption~\ref{ass:map} may be chosen as \(\alpha_J(r)=mr\). Hence \(R_\rho=\eps_{\mathrm{tot}}(\rho)/m\), and substitution into \eqref{eq:main_parameter_bound} gives \eqref{eq:strong_parameter_bound}.
\end{proof}

\begin{remark}[Interpretation and co-design of the ultimate bound]
\label{rem:rho_tradeoff}
The radius \(R_\rho\) is not an independent tuning variable. It is determined by
\[
 R_\rho=\alpha_J^{-1}\!\left(
 C_\Delta\rho+\frac{\overline\nu_1}{\rho}+h_g
 \right),
\]
so the parameter-error bound can be written as
\[
 B(\rho,\tau_{\min},h_g)
 :=\alpha_J^{-1}\!\left(
 C_\Delta\rho+\frac{\overline\nu_1}{\rho}+h_g
 \right)+\rho\tau_{\min}.
\]
The quantities \(\rho\), \(h_g\), and \(\tau_{\min}\) are design variables, but they cannot be selected independently. The speed must satisfy the local flow and self-consistency conditions, while the dwell must satisfy \(q_s<1\) and \(\tau_{\min}\ge\tau_r\). Hence, for a fixed admissible speed, the smallest feasible dwell is preferred by this bound. For a fixed feasible dwell, increasing \(\rho\) reduces the normalized sensing term but increases the dynamic-lag contribution and the factor multiplying the selected dwell; decreasing \(\rho\) has the opposite effect. In a sampled implementation, the deadband enters \(R_\rho\) directly and may be selected above an estimated sensing and numerical error floor to suppress spurious reversals.

Exact integral sliding enforces \(s=0\) and \(\dot\theta=v\); it does not enforce \(\theta=\theta^\star\). Thus the nonzero residual is not a loss of sliding accuracy. It reflects finite-rate dynamic lag, the deadband, and the excursion accumulated while a corrective reversal is delayed by the dwell requirement. For fixed \(\rho>0\), \(h_g>0\), and \(\tau_{\min}>0\), the controller generally continues to search around the optimum. In the noise-free design limit, if \(h_g\to0\), \(\rho\to0\), and \(\rho\tau_{\min}(\rho)\to0\) while all feasibility conditions remain satisfied, the guaranteed neighborhood shrinks to zero. Exact convergence with fixed parameters would require an additional zero-velocity or vanishing-speed mechanism, which is not part of the present derivative-ratio architecture.
\end{remark}

\begin{remark}[Nonvanishing motion and the role of ISM]
\label{rem:ism_role}
A prescribed nonzero denominator and a nonzero applied motion are distinct requirements. Without exact rate realization,
\[
 \frac{w}{v}
 =J'(\theta)\left(1+\frac{\dot\theta-v}{v}\right)+\frac{r_y}{v}.
\]
If \(\dot\theta=0\), the gradient component disappears even though \(v\ne0\); if \(v\dot\theta<0\), its sign is reversed. Normalizing by the applied rate leads instead, for \(\varepsilon>0\), to the regularized ratio
\begin{equation}
 \widehat g_\varepsilon
 :=\frac{w\dot\theta}{\dot\theta^2+\varepsilon^2}
 =J'(\theta)\frac{\dot\theta^2}{\dot\theta^2+\varepsilon^2}
 +r_y\frac{\dot\theta}{\dot\theta^2+\varepsilon^2}.
 \label{eq:regularized_ratio_analysis}
\end{equation}
Regularization prevents numerical divergence, but the gradient multiplier tends to zero as \(\dot\theta\to0\). Its residual-channel gain also satisfies
\[
 \max_{q\in\R}\frac{|q|}{q^2+\varepsilon^2}=\frac{1}{2\varepsilon},
\]
which exposes the bias--disturbance trade-off. The integral-sliding layer removes both motion ambiguities by enforcing \(\dot\theta=v\in\{-\rho,+\rho\}\) after finite time. When \(d_\theta\equiv0\), a nominal fixed-rate integrator has the same property; the role of ISM is to preserve it under matched rate-channel perturbations. This observation applies specifically to derivative-ratio ESC; nonsmooth methods that avoid rate normalization have different information requirements.
\end{remark}

\begin{remark}[Performance with direct parameter feedthrough]
\label{rem:direct_feedthrough}
The state-dependent output \(y=h(x)+\nu\) preserves continuity of \(\dot y\) at an instantaneous reversal. For a performance output \(h(x,\theta)+\nu\), a smooth velocity reversal or an explicit differentiator reacquisition interval would provide a natural extension that accounts for the direct parameter term.
\end{remark}

\section{Tubular-reactor comparative study}
\label{sec:numerics}

\pgfplotstableread{
theta bout
0.05 0.68536025
0.069491525 0.76599639
0.088983051 0.8154684
0.10847458 0.84870071
0.1279661 0.87222055
0.14745763 0.88924622
0.16694915 0.90152349
0.18644068 0.91009396
0.2059322 0.91564412
0.22542373 0.918671
0.24491525 0.91956266
0.26440678 0.91863817
0.28389831 0.91616818
0.30338983 0.91238606
0.32288136 0.90749455
0.34237288 0.90167022
0.36186441 0.8950668
0.38135593 0.88781796
0.40084746 0.88003963
0.42033898 0.87183222
0.43983051 0.86328242
0.45932203 0.85446498
0.47881356 0.84544416
0.49830508 0.83627511
0.51779661 0.82700508
0.53728814 0.81767441
0.55677966 0.80831749
0.57627119 0.79896355
0.59576271 0.78963734
0.61525424 0.78035977
0.63474576 0.77114839
0.65423729 0.76201788
0.67372881 0.75298043
0.69322034 0.7440461
0.71271186 0.73522309
0.73220339 0.72651801
0.75169492 0.7179361
0.77118644 0.70948142
0.79067797 0.70115703
0.81016949 0.6929651
0.82966102 0.68490709
0.84915254 0.67698378
0.86864407 0.66919545
0.88813559 0.66154188
0.90762712 0.6540225
0.92711864 0.64663637
0.94661017 0.6393823
0.96610169 0.63225886
0.98559322 0.62526442
1.0050847 0.61839719
1.0245763 0.61165526
1.0440678 0.60503663
1.0635593 0.59853919
1.0830508 0.59216078
1.1025424 0.5858992
1.1220339 0.57975221
1.1415254 0.57371756
1.1610169 0.56779299
1.1805085 0.56197623
1.2 0.55626502
}\equilibdata

\pgfplotstableread{
t prop_theta pesc_theta stw_theta prop_b pesc_b stw_b
0 1 1 1 0.62017648 0.62017648 0.62017648
1 1.0096015 1.0174122 1.0071776 0.61827317 0.61699333 0.61890391
2 0.9988133 1.0290097 1.0143551 0.61872892 0.61183719 0.61640456
3 0.98802514 1.030558 1.0215327 0.62248274 0.60954711 0.6139125
4 0.97723697 1.0210194 1.0287103 0.62628928 0.61099092 0.61143748
5 0.96644881 1.0032591 1.0358878 0.63013402 0.61582467 0.60897939
6 0.95566065 0.98299129 1.0430654 0.63401723 0.62260068 0.60653813
7 0.94487248 0.96641207 1.050243 0.63793917 0.62926699 0.60411358
8 0.93408432 0.95790762 1.0574205 0.64190008 0.6338941 0.60170563
9 0.92329615 0.9588659 1.0645981 0.64590019 0.63532501 0.59931417
10 0.91250799 0.9676372 1.0717757 0.64993969 0.63352105 0.59693909
11 0.90171983 0.98022549 1.0663925 0.65401876 0.62952526 0.59653622
12 0.89093166 0.99138254 1.0592149 0.65813756 0.6250955 0.59885859
13 0.8801435 0.99603811 1.0520374 0.66229622 0.62214859 0.60123201
14 0.86935533 0.99101724 1.0448598 0.66649483 0.62215894 0.60362149
15 0.85856717 0.97645667 1.0376822 0.67073346 0.62565623 0.60602714
16 0.847779 0.95593971 1.0305047 0.67501213 0.63201476 0.60844905
17 0.83699084 0.9350322 1.0233271 0.67933085 0.63966182 0.61088734
18 0.82620268 0.91901288 1.0161495 0.68368955 0.64664086 0.61334209
19 0.81541451 0.91099327 1.008972 0.68808814 0.65131075 0.61581341
20 0.80462635 0.91118701 1.0017944 0.69252646 0.65290302 0.61830139
21 0.79383818 0.91727516 0.99382225 0.69700432 0.65171824 0.62091482
22 0.78305002 0.92527021 0.98567768 0.70152146 0.6489478 0.62377876
23 0.77226186 0.93054201 0.97743146 0.70607753 0.64627564 0.62667124
24 0.76147369 0.9291344 0.96895345 0.71067215 0.64537421 0.62965693
25 0.75068553 0.91923339 0.96037262 0.71530484 0.64736599 0.63272586
26 0.73989736 0.90206426 0.95175171 0.71997503 0.65241635 0.63584356
27 0.7291092 0.88152957 0.94301729 0.72468209 0.65964524 0.63900449
28 0.71832104 0.86257822 0.93384185 0.72942527 0.66743186 0.64229095
29 0.70753287 0.84913745 0.9241696 0.7342037 0.67402827 0.64578729
30 0.69674471 0.84278412 0.91437516 0.73901643 0.6782179 0.64944272
31 0.68595654 0.84264482 0.90473576 0.74386237 0.67969998 0.65311828
32 0.67516838 0.84604518 0.89519537 0.74874029 0.67908618 0.65676704
33 0.66438022 0.8493574 0.88560009 0.75364882 0.67763121 0.66044363
34 0.65359205 0.84897493 0.8759821 0.75858643 0.67683423 0.66416775
35 0.64280389 0.84242731 0.86645379 0.76355142 0.6779823 0.66791544
36 0.63201572 0.82937249 0.85696372 0.76854192 0.681726 0.67166029
37 0.62122756 0.81193275 0.84726637 0.77355585 0.68781793 0.6754616
38 0.61043939 0.79389821 0.83733656 0.77859091 0.69516153 0.67939201
39 0.59965123 0.77905511 0.82739164 0.78364458 0.70222817 0.68341557
40 0.58886307 0.76967331 0.81768342 0.78871409 0.70766719 0.68743194
41 0.5780749 0.76593503 0.80814079 0.79379639 0.71079077 0.69138916
42 0.56728674 0.76622267 0.79852557 0.79888815 0.71174206 0.69535054
43 0.55649857 0.76781768 0.78876535 0.80398573 0.71136461 0.69939013
44 0.54571041 0.76770327 0.77894214 0.80908515 0.7108883 0.70350803
45 0.53492225 0.76338603 0.76903869 0.81418205 0.71155091 0.70768288
46 0.52413408 0.75374833 0.75885871 0.81927172 0.71422341 0.71195342
47 0.51334592 0.73965982 0.74830391 0.824349 0.71909099 0.71639474
48 0.50255775 0.72377987 0.73761842 0.82940828 0.72553015 0.72099881
49 0.49176959 0.70942486 0.72708494 0.83444349 0.73232364 0.72564761
50 0.48098143 0.69914875 0.71667516 0.83944802 0.73814753 0.7302644
51 0.47019326 0.69382123 0.70615414 0.84441472 0.74208621 0.73489546
52 0.4594051 0.69254122 0.69543049 0.84933584 0.74394525 0.7396253
53 0.44861693 0.69321091 0.68462344 0.854203 0.74425431 0.74445769
54 0.43782877 0.69329841 0.67379736 0.85900714 0.74401261 0.74934951
55 0.4270406 0.69050824 0.66289474 0.86373849 0.74434511 0.75428641
56 0.41625244 0.68344188 0.65180973 0.86838653 0.74617716 0.7593049
57 0.40546428 0.67219128 0.64074288 0.87293989 0.74994016 0.76440784
58 0.39467611 0.65846725 0.62994392 0.87738639 0.75538507 0.76949695
59 0.38388795 0.64501135 0.61934485 0.88171292 0.76163269 0.77449886
60 0.37309978 0.63448539 0.60874306 0.88590545 0.76748362 0.77945119
61 0.36231162 0.62836859 0.59798649 0.88994893 0.77187552 0.78444169
62 0.35152346 0.62647322 0.58716383 0.89382729 0.77429151 0.78950218
63 0.34073529 0.62725874 0.57631724 0.89752341 0.77492159 0.79460311
64 0.32994713 0.62851107 0.5653965 0.90101906 0.77452784 0.79972834
65 0.31915896 0.6279716 0.55429847 0.90429487 0.77414994 0.80490635
66 0.3083708 0.62391086 0.54313968 0.90733036 0.7747934 0.81014929
67 0.29758264 0.61567742 0.53212291 0.91010386 0.77714249 0.8153991
68 0.28679447 0.60405686 0.52124791 0.91259261 0.78133386 0.82058487
69 0.27600631 0.59118938 0.51029379 0.91477271 0.78686551 0.82571323
70 0.26521814 0.5798345 0.499111 0.91661918 0.79271946 0.83085991
71 0.25442998 0.57217971 0.48781919 0.91810606 0.79772128 0.83606187
72 0.24364182 0.56892685 0.4765679 0.91920643 0.80100478 0.84127007
73 0.23285365 0.56919872 0.46533745 0.91989256 0.80233798 0.84642985
74 0.22972508 0.57108002 0.45411418 0.92012723 0.8021616 0.85152658
75 0.22918568 0.57232421 0.44297675 0.91995159 0.80137868 0.85655148
76 0.23360882 0.57095779 0.43215099 0.91961172 0.80103236 0.86145565
77 0.22681228 0.56573935 0.42170716 0.91925922 0.80200028 0.86616544
78 0.23209848 0.5565634 0.41138062 0.91911086 0.8047613 0.87065799
79 0.23048026 0.544727 0.40095911 0.91907789 0.80923194 0.87499668
80 0.22864627 0.53266145 0.39054953 0.91910102 0.81472984 0.87923603
81 0.23414823 0.52298541 0.38039097 0.91908094 0.82017275 0.88334884
82 0.22735169 0.51739671 0.37052669 0.91908784 0.82447931 0.88726728
83 0.23177483 0.51606138 0.36079188 0.91909763 0.82697026 0.89096433
84 0.23134331 0.51773078 0.3512084 0.91908101 0.82759096 0.89445491
85 0.22778321 0.52036625 0.34200159 0.91910276 0.82689086 0.89773286
86 0.2349034 0.52181511 0.33322859 0.91908941 0.82579694 0.90076056
87 0.2278911 0.5202795 0.32461145 0.91909471 0.8253097 0.90352455
88 0.23867926 0.51473855 0.31588457 0.91911888 0.82623583 0.90606483
89 0.24180783 0.50540743 0.30716034 0.9191909 0.82896099 0.9084267
90 0.24213147 0.49389143 0.29880909 0.91923225 0.83328691 0.91060235
91 0.23813985 0.48271874 0.29098378 0.91924066 0.83845243 0.91254311
92 0.2454758 0.47436679 0.28356589 0.91943572 0.84338995 0.91421669
93 0.23867926 0.47024401 0.27652003 0.91953228 0.84711067 0.91562883
94 0.24180783 0.47019678 0.27014399 0.91954035 0.84904823 0.91679577
95 0.24267088 0.47277557 0.26470847 0.91953471 0.84922923 0.91772088
96 0.23781621 0.47589092 0.25991603 0.91954786 0.84822747 0.91841435
97 0.24515216 0.47740444 0.25522567 0.91953743 0.8469543 0.91891717
98 0.23878714 0.47561365 0.25068631 0.91954408 0.84638191 0.9192815
99 0.24116054 0.4697142 0.24738547 0.91954727 0.84726548 0.91953732
100 0.24277876 0.46014835 0.2473454 0.91953567 0.84991059 0.9196619
101 0.23749256 0.4486533 0.24730533 0.91954739 0.85405545 0.9196738
102 0.24428911 0.43778878 0.24726526 0.91953413 0.85893447 0.91961692
103 0.23965019 0.42995245 0.24722519 0.91953699 0.86353664 0.91955937
104 0.24029748 0.42640478 0.24718512 0.91954325 0.86696294 0.91954511
105 0.24364182 0.42688491 0.24714505 0.91953724 0.86871427 0.91954468
106 0.23662951 0.42987045 0.24710498 0.91954572 0.86881663 0.91954523
107 0.23954231 0.4331836 0.24706491 0.91955328 0.8677946 0.91954579
108 0.2331773 0.43465041 0.24702484 0.91955358 0.86651253 0.91954634
109 0.23997384 0.43262049 0.24698477 0.91956579 0.86590865 0.91954688
110 0.23641375 0.42637504 0.2469447 0.91948025 0.86670484 0.91954742
111 0.23652163 0.41648744 0.24690463 0.91945574 0.86917602 0.91954795
112 0.23986596 0.40486167 0.24686455 0.91944133 0.87303873 0.91954847
113 0.23296153 0.39413463 0.24682448 0.91945132 0.87753706 0.91954899
114 0.23997384 0.38664472 0.24678441 0.91944754 0.88172922 0.9195495
115 0.23641375 0.3835212 0.24674434 0.91944246 0.88482563 0.91955
116 0.23652163 0.38433403 0.24670427 0.91945199 0.88641533 0.91955049
117 0.24029748 0.38741925 0.2466642 0.91943844 0.88654363 0.91955097
118 0.23350094 0.39057202 0.24662413 0.9194527 0.88567905 0.91955145
119 0.23943443 0.39165544 0.24658406 0.91944832 0.88458511 0.91955192
120 0.23716892 0.38905465 0.24654399 0.91944615 0.88410286 0.91955238
121 0.23555069 0.38214231 0.24650392 0.91945878 0.88488708 0.91955284
122 0.24105266 0.37163871 0.24646385 0.91944581 0.88716908 0.91955329
123 0.23404035 0.35955333 0.24642378 0.91945289 0.89065405 0.91955373
124 0.23889502 0.34856358 0.24638371 0.91945047 0.8946385 0.91955416
125 0.23749256 0.34099595 0.24634364 0.91944118 0.89830256 0.91955458
126 0.23522705 0.33792003 0.24630357 0.91945387 0.90100856 0.919555
127 0.24116054 0.3388702 0.2462635 0.91944052 0.90245522 0.91955541
128 0.23414823 0.34218 0.24622343 0.9194492 0.90268388 0.91955581
129 0.23857138 0.34555844 0.24618336 0.91944946 0.90203713 0.91955621
130 0.23760044 0.34671309 0.24614329 0.91943818 0.90110924 0.91955659
131 0.23511917 0.34396628 0.24610322 0.91945031 0.90061634 0.91955697
132 0.24148418 0.33676672 0.24606315 0.91943764 0.9011462 0.91955734
133 0.23414823 0.32598785 0.24602308 0.91944556 0.90289783 0.91955771
134 0.23878714 0.31380252 0.24598301 0.91944846 0.90558491 0.91955806
135 0.23803197 0.30295782 0.24594294 0.91944084 0.90858416 0.91955841
136 0.23479552 0.29574171 0.24590287 0.91945705 0.91123932 0.91955875
137 0.24169995 0.29321089 0.2458628 0.91944217 0.91311582 0.91955909
138 0.23468764 0.2949298 0.24582273 0.91945132 0.91406983 0.91955941
139 0.23824773 0.29917343 0.24578266 0.91945467 0.91417486 0.91955973
140 0.23857138 0.30349438 0.24574259 0.91944349 0.91364865 0.91956004
141 0.23425611 0.3054366 0.24570252 0.91945728 0.91285679 0.91956035
142 0.24116054 0.30318959 0.24566245 0.91944349 0.91230031 0.91956064
143 0.23565857 0.29614397 0.24562238 0.91945005 0.91245005 0.91956093
144 0.23749256 0.28521128 0.24558231 0.91945584 0.91348663 0.91956121
145 0.23965019 0.27263831 0.24554224 0.9194437 0.91517367 0.91956148
146 0.23350094 0.26130024 0.24550217 0.91945766 0.91699491 0.91956175
147 0.24029748 0.2537439 0.2454621 0.91945076 0.91845201 0.91956201
148 0.23630586 0.25132323 0.24542203 0.91945253 0.91930067 0.91956226
149 0.23684527 0.25376596 0.24538196 0.91946218 0.91958268 0.9195625
150 0.2401896 0.25931018 0.24534189 0.91944875 0.9194831 0.91956273
151 0.23296153 0.26525058 0.24530182 0.91945817 0.91916514 0.91956296
152 0.24029748 0.26871782 0.24526175 0.91945112 0.91874133 0.91956318
153 0.23673739 0.26755314 0.24522168 0.91945016 0.918384 0.91956339
154 0.23630586 0.2610079 0.24518161 0.91946162 0.91834762 0.9195636
155 0.24072901 0.25003539 0.24514154 0.91944365 0.91876496 0.91956379
156 0.23360882 0.23709051 0.24510147 0.91945819 0.91945542 0.91956398
157 0.23975808 0.22538906 0.2450614 0.919455 0.92001094 0.91956416
158 0.23706104 0.21782245 0.24502133 0.91944899 0.9200902 0.91956434
159 0.2360901 0.21602841 0.24498126 0.91946205 0.91965437 0.9195645
160 0.24105266 0.21992334 0.24494119 0.91944707 0.91897639 0.91956466
161 0.23425611 0.2276799 0.24490112 0.91945861 0.9184486 0.91956481
162 0.23889502 0.23618594 0.24486105 0.91945529 0.91831627 0.91956496
163 0.23813985 0.24198912 0.24482098 0.91944998 0.9185189 0.91956509
164 0.2349034 0.24242123 0.24478091 0.91946293 0.91882216 0.91956522
165 0.24213147 0.23650232 0.24474084 0.91945076 0.91910711 0.91956534
166 0.23479552 0.22533994 0.24470077 0.91945613 0.91937123 0.91956545
167 0.23878714 0.21181601 0.2446607 0.91945839 0.91947497 0.91956556
168 0.23867926 0.1996322 0.24462063 0.91944806 0.91911446 0.91956566
169 0.23479552 0.19214295 0.24458056 0.91946644 0.91809233 0.91956575
170 0.24213147 0.19135222 0.24454049 0.91945341 0.91656029 0.91956583
171 0.23511917 0.19725785 0.24450042 0.91946138 0.91501651 0.9195659
172 0.23824773 0.20773647 0.24446035 0.91946561 0.91408364 0.91956597
173 0.2390029 0.21909603 0.24442028 0.91945242 0.91417615 0.91956603
174 0.23457976 0.22723269 0.24438021 0.91946966 0.91518878 0.91956608
175 0.24245512 0.22908339 0.24434014 0.91945572 0.9165162 0.91956612
176 0.23781621 0.22374078 0.24430007 0.91945537 0.91757615 0.91956616
177 0.24493639 0.21269647 0.24426 0.91945814 0.9181602 0.91956619
178 0.23954231 0.19927511 0.24421993 0.91943985 0.91817547 0.91956621
179 0.24116054 0.18759616 0.24417986 0.91951301 0.91745369 0.91956622
180 0.24342605 0.18132956 0.24413979 0.91953157 0.91595719 0.91956622
}\representativedata

\pgfplotstableread{
p proposed stw_gradient pesc_tuned pesc_fixed
0.005 3.1591812e-06 1e-09 0.00066410996 1.8320251e-05
0.015 3.894452e-06 1e-09 0.00067712954 0.00014644509
0.025 8.0331555e-06 1e-09 0.000688843 0.00022946104
0.035 1.0768605e-05 1e-09 0.00069928284 0.00047620496
0.045 1.6184824e-05 4.3721941e-09 0.00070339761 0.00060758121
0.055 1.685172e-05 1.1548313e-07 0.00071363847 0.00081558238
0.065 1.9844844e-05 1.0900421e-06 0.00071556288 0.0011306665
0.075 2.4949114e-05 1.8012338e-06 0.0007216842 0.0012978119
0.085 3.3361484e-05 2.0607766e-06 0.00072700135 0.0023586226
0.095 3.620028e-05 2.6392977e-06 0.00073165581 0.0026357853
0.105 4.0189717e-05 3.3695054e-06 0.00073712527 0.0037015558
0.115 4.6625611e-05 5.1408359e-06 0.00074113165 0.004037099
0.125 4.7026517e-05 5.5953882e-06 0.00074841907 0.0045364323
0.135 5.0520232e-05 6.5183793e-06 0.00075837914 0.0049766499
0.145 5.2613463e-05 7.6531516e-06 0.00076627663 0.00562145
0.155 5.3788145e-05 8.1566339e-06 0.00076734888 0.0063258218
0.165 6.2732797e-05 9.1978772e-06 0.00076949502 0.0084721436
0.175 6.5427358e-05 1.0562928e-05 0.00077388307 0.0091447397
0.185 6.6431641e-05 1.1597719e-05 0.0007758805 0.01156428
0.195 6.7888912e-05 1.2404565e-05 0.00077779956 0.013289977
0.205 7.3021855e-05 1.3360763e-05 0.00078169476 0.014397068
0.215 7.5059156e-05 1.5074603e-05 0.0007874481 0.016447938
0.225 7.6601726e-05 1.5578515e-05 0.00079443978 0.017246593
0.235 8.0442669e-05 2.0974858e-05 0.00080175918 0.018552102
0.245 8.1733082e-05 2.3352315e-05 0.00080480023 0.019270033
0.255 8.7282988e-05 2.5134318e-05 0.00080598475 0.019498556
0.265 8.7754715e-05 2.6677322e-05 0.00080912612 0.021816546
0.275 9.108846e-05 2.7497963e-05 0.0008183594 0.025103324
0.285 9.3203682e-05 2.8302041e-05 0.00082440442 0.030135459
0.295 9.6891762e-05 2.9403538e-05 0.00083270828 0.03300804
0.305 9.8999494e-05 3.2798789e-05 0.00083625066 0.035971873
0.315 0.00010258881 3.4661494e-05 0.00084313541 0.03852227
0.325 0.00010444682 3.6700001e-05 0.00084609931 0.0411834
0.335 0.00010727855 3.7760245e-05 0.00085563182 0.042974817
0.345 0.00010869398 4.3398358e-05 0.00085953314 0.043213761
0.355 0.00011094925 4.9735959e-05 0.00086650685 0.043769014
0.365 0.00011417314 5.1565266e-05 0.00086990068 0.052359351
0.375 0.00011547156 5.3430019e-05 0.00087528374 0.056410928
0.385 0.00011974267 5.682715e-05 0.0008800349 0.059548566
0.395 0.00012324818 6.2566977e-05 0.0008852731 0.064840177
0.405 0.00012396509 6.6737844e-05 0.00089043411 0.068066111
0.415 0.00012572514 6.8708698e-05 0.00089548512 0.072626337
0.425 0.00012986134 7.3961201e-05 0.00090168565 0.073676997
0.435 0.00013274663 7.5436924e-05 0.00090514299 0.075686489
0.445 0.00013863596 7.7692842e-05 0.00090979912 0.0789608
0.455 0.0001404057 8.1383365e-05 0.0009156694 0.081945583
0.465 0.00014299341 8.2303418e-05 0.00092694006 0.089947978
0.475 0.00014570509 8.8661129e-05 0.00093813792 0.09152567
0.485 0.00014746444 9.0340516e-05 0.00094003463 0.093318968
0.495 0.00014791562 9.2021343e-05 0.00094860415 0.097555316
0.505 0.00015165009 0.00010510953 0.00096103268 0.10626314
0.515 0.00015528515 0.00011710901 0.0009787811 0.1125855
0.525 0.00015704311 0.00012500165 0.00098421255 0.11535877
0.535 0.00015838937 0.00012973854 0.00099718662 0.11955442
0.545 0.00016063729 0.00013124076 0.0010093882 0.12488544
0.555 0.00016451851 0.00013293349 0.001024266 0.13298485
0.565 0.00016669529 0.00013989381 0.0010410993 0.13410487
0.575 0.00017033693 0.00014454068 0.0010443103 0.13743844
0.585 0.00017233384 0.00015514632 0.0010487759 0.13943523
0.595 0.00017316269 0.00016285094 0.0010679368 0.14183565
0.605 0.00017882114 0.00017522249 0.0010836403 0.14412682
0.615 0.00018542929 0.00017769639 0.0010930863 0.1455978
0.625 0.00018627883 0.0001893958 0.0010995127 0.14646585
0.635 0.00018729469 0.00020407782 0.00110958 0.14792404
0.645 0.00019452566 0.0002085356 0.0011378389 0.1583478
0.655 0.00019681519 0.00021411872 0.001158985 0.16273481
0.665 0.00019939836 0.00022759567 0.0011858914 0.16684557
0.675 0.00020259832 0.00023784483 0.0012205048 0.17707538
0.685 0.0002100985 0.0002512174 0.00122665 0.18294528
0.695 0.00021138158 0.00026639826 0.0012378075 0.18650421
0.705 0.00021808436 0.0002879999 0.001256267 0.19076277
0.715 0.00022319959 0.00029813896 0.0013073454 0.19563419
0.725 0.00022627322 0.00031226421 0.0013372366 0.19877478
0.735 0.00022913576 0.00033199589 0.0013622986 0.20108516
0.745 0.00023543857 0.00035735815 0.0013817389 0.20463711
0.755 0.00023818505 0.00037796874 0.0014639865 0.20850777
0.765 0.00024222107 0.00038394271 0.0015542496 0.21144441
0.775 0.00025091981 0.0004110944 0.0016172461 0.21480144
0.785 0.000252871 0.00046187024 0.0016673852 0.21878364
0.795 0.00026207651 0.00053283245 0.001751629 0.22431599
0.805 0.00026735155 0.00059303804 0.0018396825 0.22654782
0.815 0.00028177497 0.00066339815 0.0019156926 0.23205656
0.825 0.00028512846 0.00081422393 0.0019309405 0.23237083
0.835 0.00028981257 0.001041868 0.0019847673 0.23840518
0.845 0.0002950977 0.0011098764 0.0020633651 0.2459719
0.855 0.00030155452 0.002288325 0.0022138844 0.24731395
0.865 0.00030769607 0.0047356565 0.0023717316 0.25258116
0.875 0.00031204824 0.0063221639 0.0024485939 0.25680702
0.885 0.00031597417 0.0080352165 0.0025409494 0.25840511
0.895 0.00032596295 0.0098812795 0.0025987611 0.26501834
0.905 0.00033293358 0.013074565 0.0026785382 0.26783746
0.915 0.00033679561 0.015603338 0.0027240737 0.26925648
0.925 0.00034985464 0.021221403 0.0027457323 0.27125538
0.935 0.00035263681 0.038361511 0.0028416614 0.27594244
0.945 0.00036323772 0.045533642 0.0032831045 0.27721452
0.955 0.00038297188 0.060867013 0.0033591259 0.27953693
0.965 0.00040945713 0.068042361 0.0034234041 0.28127036
0.975 0.00041948148 0.082664353 0.0036443354 0.293149
0.985 0.00044481232 0.091732759 0.024581738 0.29770536
0.995 0.00056373498 0.10723088 0.073279077 0.30714117
}\heldoutcdfdata

\pgfplotstableread{
kappa prop_success noism_success pesc_success stw_success prop_q90 noism_q90 pesc_q90 stw_q90 prop_stall noism_stall prop_opp noism_opp prop_mismatch noism_mismatch stw_att_nz stw_beta_nz noism_stall_pct noism_opp_pct
0.0 1 1 1 0.85 0.00035099068 0.00035099068 0.0022503807 0.026696131 0 0 0 0 0 0 0.28905114 0.54840328 0 0
0.5 1 1 1 1 0.00033189255 0.0031526119 0.002248757 0.0011153174 0 0 0 0 0.00032241971 0.22204823 0.78120775 0.25064979 0 0
0.9 1 0.31 1 1 0.00033082135 0.0069399972 0.0022474635 0.0034956072 0 0 0 0 0.00058035548 0.39968681 0.66180175 0.3424788 0 0
1.1 1 0 1 0.96 0.00032601196 0.0091998859 0.0022468185 0.0046876358 0 0.040880152 0 0.019406903 0.00070932337 0.48850611 0.61052242 0.38467106 4.0880152 1.9406903
1.3 1 0 1 0.57 0.0003576017 0.013418192 0.0022461747 0.0061586299 0 0.040419412 0 0.052203172 0.00083829125 0.5773254 0.56149933 0.42384048 4.0419412 5.2203172
}\ratesweepdata

\subsection{Reactor model and equilibrium optimum}
We use the tubular plug-flow reactor benchmark of \citet{TrollbergJacobsen2018}. It considers the irreversible consecutive reaction
\[
 \mathrm A\xrightarrow{k_1}\mathrm B\xrightarrow{k_2}\mathrm C,
\]
where \(\mathrm A\) is the feed reactant, \(\mathrm B\) is the desired intermediate product, and \(\mathrm C\) is an undesired downstream product. The normalized axial feed velocity \(\theta\in\Theta=[0.05,1.20]\) sets the residence time. A large velocity gives insufficient conversion of \(\mathrm A\) into \(\mathrm B\), whereas a small velocity promotes further conversion of \(\mathrm B\) into \(\mathrm C\); hence the steady-state outlet concentration of \(\mathrm B\) has an interior maximum. Because \(\mathrm C\) neither feeds back into the \(\mathrm A\)--\(\mathrm B\) balances nor enters the performance output, only the concentrations of \(\mathrm A\) and \(\mathrm B\) are retained as states.

Let \(a(t,z)\) and \(b(t,z)\) denote these concentrations at normalized axial position \(z\in[0,1]\), with \(z=0\) at the inlet and \(z=1\) at the outlet. Their convection--reaction balances are
\begin{align}
 \partial_t a&=-\theta\,\partial_z a-k_1a,\nonumber\\
 \partial_t b&=-\theta\,\partial_z b+k_1a-k_2b,
 \qquad z\in[0,1],
 \label{eq:reactor_pde}
\end{align}
with \(a(t,0)=1\) and \(b(t,0)=0\).

Set \(\Delta z:=1/N\), \(N=40\), and let \(a_i,b_i\), \(i=1,\ldots,N\), denote the upwind cell states, with inlet values \(a_0=1\), \(b_0=0\). The resulting 80-state model is
\begin{align}
 \dot a_i&=-\frac{\theta}{\Delta z}(a_i-a_{i-1})-k_1a_i,\nonumber\\
 \dot b_i&=-\frac{\theta}{\Delta z}(b_i-b_{i-1})+k_1a_i-k_2b_i,
 \quad i=1,\ldots,N.
 \label{eq:reactor_ode}
\end{align}
Each run starts from the discrete steady state associated with its sampled \(\theta(0),k_1,k_2\). The state \(b_N\) approximates the outlet concentration \(b(t,1)\); denote the resulting discrete equilibrium outlet map by \(b_N^{\rm ss}(\theta)\). For each kinetic realization, the equilibrium performance map in \eqref{eq:J_def} is \(J(\theta):=-b_N^{\rm ss}(\theta)\). The measured minimization output is
\begin{equation}
 y(t)=-b_N(t)+\nu(t).
 \label{eq:reactor_output}
\end{equation}
At steady state without the sensing disturbance, minimizing \(y\) is equivalent to maximizing the outlet concentration of the desired intermediate product \(\mathrm B\). The objective is outlet concentration, not material throughput \(\theta b_N\). For each kinetic realization, \(\theta^\star\) is obtained by bounded scalar minimization of the discrete steady-state performance map over \(\Theta\), and \(b_N^\star\) is evaluated at the resulting optimizer. Both quantities depend on \((k_1,k_2)\) and are used only for evaluation. For the mechanism diagnostics, \(J'(\theta)\) is evaluated offline from the same steady-state recursion and is not supplied to any controller.

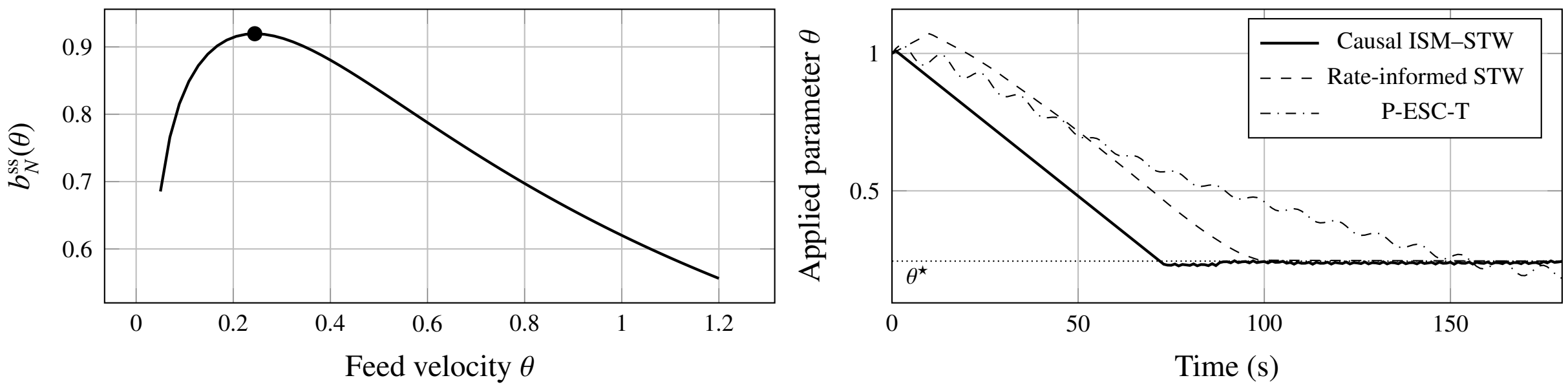
\begin{figure*}[t]
\centering
\begin{tikzpicture}
\begin{groupplot}[
 group style={group size=2 by 1,horizontal sep=1.2cm},
 width=0.46\textwidth,height=0.25\textwidth,
 grid=major,legend style={font=\scriptsize},tick label style={font=\scriptsize},label style={font=\small}]
\nextgroupplot[xlabel={Feed velocity $\theta$},ylabel={$b_N^{\rm ss}(\theta)$}]
 \addplot[black,thick] table[x=theta,y=bout]{\equilibdata};
 \addplot[only marks,mark=*,black] coordinates {(0.2442173,0.9195638)};
\nextgroupplot[xlabel={Time (s)},ylabel={Applied parameter $\theta$},xmin=0,xmax=180,enlarge x limits=false,legend pos=north east]
 \addplot[black,thick] table[x=t,y=prop_theta]{\representativedata};\addlegendentry{Causal ISM--STW}
 \addplot[black,dashed] table[x=t,y=stw_theta]{\representativedata};\addlegendentry{Rate-informed STW}
 \addplot[black,dashdotted] table[x=t,y=pesc_theta]{\representativedata};\addlegendentry{P-ESC-T}
 \addplot[black,densely dotted,thin,forget plot] coordinates {(0,0.2442173) (180,0.2442173)}
  node[pos=0.015,below right,font=\scriptsize,inner sep=1pt] {$\theta^\star$};
\end{groupplot}
\end{tikzpicture}
\caption{Nominal equilibrium map (left) and representative parameter trajectories (right). The marker and horizontal dotted line denote the discretized nominal optimum.}
\label{fig:map_trace}
\end{figure*}

Figure~\ref{fig:map_trace} shows the nominal map and a representative case with \(\theta(0)=1\), \((k_1,k_2)=(1,0.02)\), sensing phases \((0.3,1.4,2.5)\), and no rate disturbance. This benchmark is challenging for dynamic ESC because changing \(\theta\) changes both the equilibrium outlet map and the convective transport time. Consequently, the plant bandwidth and input-output phase lag vary with the operating point, so one periodic probing frequency need not produce the same demodulation phase over the full search interval. Continuation studies of this model report multiple periodic stationary solutions under classical ESC \citep{TrollbergJacobsen2018}.

\subsection{Common numerical safeguards and compared methods}
All plant and continuous controller states are advanced by forward Euler with the common step \(\Delta t\). Let \(N_s\) denote the number of update intervals, so the simulated horizon is \(T=N_s\Delta t\), and define
\[
 t_j^{\rm s}:=j\Delta t,\qquad j=0,1,\ldots,N_s.
\]
At sampling instant \(t_j^{\rm s}\), a controller may generate a raw rate command \(q_{c,j}\). After the matched perturbation is included and the common safeguards are applied, the resulting rate \(q_{a,j}\) is held on \([t_j^{\rm s},t_{j+1}^{\rm s})\) for \(j=0,\ldots,N_s-1\). We write \(\theta_j:=\theta(t_j^{\rm s})\), \(\xi_j:=\xi(t_j^{\rm s})\), and \(y_j:=y(t_j^{\rm s})\) for \(j=0,\ldots,N_s\), and \(d_{\theta,j}:=d_\theta(t_j^{\rm s})\) for the update intervals. For a relay-based realization, \(v_j^+\) denotes the prescribed search velocity selected at \(t_j^{\rm s}\) and held on that interval. Let \(\widetilde q_{a,j}\) denote the complete pre-safeguard rate candidate, including the matched perturbation exactly once. Define
\begin{align}
 \operatorname{sat}_{v_{\max}}(q)&:=\min\{\max\{q,-v_{\max}\},v_{\max}\},\nonumber\\
 \mathcal S_{\Theta,\Delta t}(\theta,q)&:=\min\!\left\{\max\!\left\{q,
 \frac{\underline\theta-\theta}{\Delta t}\right\},
 \frac{\overline\theta-\theta}{\Delta t}\right\},
 \label{eq:common_safeguard}
\end{align}
and apply
\begin{equation}
 q_{a,j}=\mathcal S_{\Theta,\Delta t}\!\left(
 \theta_j,\operatorname{sat}_{v_{\max}}(\widetilde q_{a,j})\right),
 \qquad
 \theta_{j+1}=\theta_j+\Delta t\,q_{a,j},
 \label{eq:common_numeric_channel}
\end{equation}
with \(v_{\max}=0.10\) and the reactor interval \(\Theta\) defined above. Thus \(\mathcal S_{\Theta,\Delta t}\) keeps \(\theta_{j+1}\in\Theta\). For the smooth comparison controllers, \(\widetilde q_{a,j}=q_{c,j}+d_{\theta,j}\). Let \(\#\mathcal A\) denote the number of elements in a finite sample set \(\mathcal A\), set \(\mathcal J_s:=\{0,\ldots,N_s-1\}\), and define \(\bar q_j:=\operatorname{sat}_{v_{\max}}(\widetilde q_{a,j})\). The safeguard-activation fractions are
\begin{align}
 p_{\partial\Theta}&:=\frac{1}{N_s}\#\{j\in\mathcal J_s:q_{a,j}\ne\bar q_j\},\nonumber\\
 p_{\rm cap}&:=\frac{1}{N_s}\#\{j\in\mathcal J_s:\bar q_j\ne\widetilde q_{a,j}\}.
 \label{eq:safeguard_fractions}
\end{align}
For a trajectory of duration \(T\), define
\begin{equation}
 \mathcal A_\theta:=\frac{\operatorname{TV}(\theta)}{T},
 \qquad
 \operatorname{TV}(\theta):=\sum_{j=0}^{N_s-1}|\theta_{j+1}-\theta_j|.
 \label{eq:metric_activity}
\end{equation}
In the continuous-time method descriptions and diagnostics, \(q_a(t)\) denotes the zero-order-held interpolation of \(q_{a,j}\). Table~\ref{tab:gradient_structures} summarizes the distinct information mechanisms. P-ESC supplies periodic excitation and therefore has no derivative denominator. The two derivative-ratio methods instead rely on applied parameter motion.

\begin{table*}[t]
\centering\scriptsize
\caption{Gradient-information structures of the compared methods.}
\label{tab:gradient_structures}
\begin{tabular}{@{}p{0.15\textwidth}p{0.28\textwidth}p{0.18\textwidth}p{0.28\textwidth}@{}}
\toprule
Method & Gradient information & Rate signal used in normalization & Near-zero applied rate\\
\midrule
P-ESC & Periodic perturbation and demodulation & None & No rate denominator; periodic probing remains present\\
Rate-informed STW & Regularized output-derivative ratio & Measured applied rate \(q_a\) & Gradient multiplier is attenuated toward zero\\
Causal ISM--STW & \(w/v\) with fixed \(v\in\{-\rho,+\rho\}\) & Prescribed search rate \(v\) & After continuous-time sliding capture, \(\dot\theta=v\ne0\); sampled rate mismatch is evaluated empirically\\
\bottomrule
\end{tabular}
\end{table*}

The principal comparison uses a causal sampled proximal-feedback realization. At the beginning of update \(j\), let \(v_j\in\{-\rho,+\rho\}\) be the incoming search velocity and let \(n_{d,j}\) count the elapsed samples since the most recent reversal. The measurement \(y_j\) first produces the post-update differentiator states
\begin{align}
 e_{y,j}&:=\widehat y_j-y_j,\nonumber\\
 \widehat y_j^+&:=\widehat y_j+\Delta t\{-\lambda_1|e_{y,j}|^{1/2}\sgn(e_{y,j})+w_j\},\nonumber\\
 w_j^+&:=w_j-\Delta t\,\lambda_2\sgn(e_{y,j}).
 \label{eq:sampled_stw}
\end{align}
The updated derivative estimate is used for the relay decision at the same sampling instant; no future measurement is used. Let \(N_{\min}:=\lceil\tau_{\min}/\Delta t\rceil\) and \(\widehat g_j^{\rm pre}:=w_j^+/v_j\). If
\[
 t_j^{\rm s}\ge T_A,\qquad
 n_{d,j}\ge N_{\min},\qquad
 |\widehat g_j^{\rm pre}|>h_g,\qquad
 v_j\widehat g_j^{\rm pre}>0,
\]
set
\[
 v_j^+:=-\rho\sgn(\widehat g_j^{\rm pre}),\qquad n_{d,j}^+:=0;
\]
otherwise set \(v_j^+:=v_j\) and \(n_{d,j}^+:=n_{d,j}\). The selected search velocity is held on \([t_j^{\rm s},t_{j+1}^{\rm s})\), and the stored controller states are
\begin{equation}
 v_{j+1}:=v_j^+,
 \quad
 n_{d,j+1}:=n_{d,j}^++1,
 \quad
 \widehat y_{j+1}:=\widehat y_j^+,
 \quad
 w_{j+1}:=w_j^+.
 \label{eq:sampled_relay}
\end{equation}
Define the soft-threshold map
\[
 \operatorname{soft}(r,\lambda):=\sgn(r)\max\{|r|-\lambda,0\}.
\]
With the measured sliding error \(s_j:=\theta_j-\xi_j\), set
\begin{align}
 \xi_{j+1}&=\xi_j+\Delta t\,v_j^+,\nonumber\\
 q_{c,j}&=v_j^++
 \frac{\operatorname{soft}(s_j,\Delta t\,k_I)-s_j}{\Delta t},\nonumber\\
 \widetilde q_{a,j}&=q_{c,j}+d_{\theta,j}.
 \label{eq:causal_proximal_ism}
\end{align}
The common safeguard \eqref{eq:common_numeric_channel} is then applied. When the safeguards are inactive, this update gives
\(s_{j+1}=\operatorname{soft}(s_j,\Delta t k_I)+\Delta t d_{\theta,j}\); it uses the measured \(s_j\) but not the unknown disturbance. A limited step-size sensitivity also uses
\begin{equation}
 q_{c,j}=v_j^+-k_I\sgn(s_j),
 \qquad
 \widetilde q_{a,j}=q_{c,j}+d_{\theta,j}.
 \label{eq:explicit_sampled_ism}
\end{equation}
Both updates causally approximate the continuous-time Filippov inclusion. The step-size study characterizes their numerical behavior; a sampled-data theorem requires an additional discrete-time analysis. The proposed controller uses \(\xi_0=\theta_0\), \(s_0=0\), \(v_0=+\rho\), and \(n_{d,0}=N_{\min}\). The differentiator starts from \(\widehat y_0=y_0\) and \(w_0=0\).

The No-ISM ablation retains this differentiator, relay, deadband, dwell logic, activation time, initialization, selected gains, and common safeguards, but removes the proximal integral-sliding correction. Its complete pre-safeguard rate is
\[
 \widetilde q_{a,j}=v_j^+ + d_{\theta,j}.
\]

For perturbation ESC, let \(r_h,r_l,\omega,k_{\rm esc},a_p>0\) and \(\phi\in\R\), and let \(\eta_h\), \(\eta_l\), and \(\bar\theta\) satisfy
\begin{align}
 \dot\eta_h&=r_h\omega(y-\eta_h),\nonumber\\
 \dot\eta_l&=r_l\omega\{(y-\eta_h)\sin(\omega t+\phi)-\eta_l\},\nonumber\\
 \dot{\bar\theta}&=-k_{\rm esc}\eta_l,
 \qquad \theta_d=\bar\theta+a_p\sin(\omega t).
 \label{eq:pesc_comparator}
\end{align}
At each sample, the filter and optimizer states are updated simultaneously from their current values using \(y_j\). The updated optimizer state defines \(\theta_d(t_{j+1}^{\rm s})=\bar\theta_{j+1}+a_p\sin(\omega t_{j+1}^{\rm s})\), which is converted into the raw rate command
\begin{equation}
 q_{c,j}=\frac{\theta_d(t_{j+1}^{\rm s})-\theta_j}{\Delta t}.
 \label{eq:pesc_raw_rate}
\end{equation}
The initial states are \(\bar\theta_0=\theta_0\), \(\eta_{h,0}=y_0\), and \(\eta_{l,0}=0\). The fixed-gain P-ESC-F uses
\[
 (\omega,a_p,k_{\rm esc},r_h,r_l,\phi)
 =(0.4,0.001,0.01,0.1,0.1,0),
\]
and P-ESC-T tunes the same six quantities.

All parameters of the rate-informed STW comparator are positive. Define \(e_c:=\widehat y_c-y\) and write
\begin{align}
 \dot{\widehat y}_c&=-\lambda_{1c}|e_c|^{1/2}\sgn(e_c)+w_c,\nonumber\\
 \dot w_c&=-\lambda_{2c}\sgn(e_c),\nonumber\\
 g_r&=\frac{w_cq_a}{q_a^2+\varepsilon_c^2},
 \qquad \dot g_l=f_g(g_r-g_l).
 \label{eq:stw_comparator_diff}
\end{align}
Let \(\sgn_{h_c}(g)=0\) for \(|g|\le h_c\) and \(\sgn_{h_c}(g)=\sgn(g)\) otherwise. During the symmetric probe,
\begin{equation}
 q_c(t)=
 \begin{cases}
 +v_p, & 0\le t<T_p/2,\\
 -v_p, & T_p/2\le t<T_p,
 \end{cases}
 \label{eq:stw_probe}
\end{equation}
while the differentiator and gradient filter continue to evolve and the optimizer state is held at \(z_c=0\). In the sampled implementation, the probe is terminated at the first sampling instant \(t_j^{\rm s}\ge T_p\), and the optimizer law
\begin{equation}
 \dot z_c=-\alpha_c\sgn_{h_c}(g_l),
 \qquad
 q_c=-\lambda_c\sqrt{|g_l|}\sgn_{h_c}(g_l)+z_c
 \label{eq:stw_comparator_optimizer}
\end{equation}
is activated. Initialize the completed-interval rate by \(q_a(0^-)=v_p\). At \(t_j^{\rm s}\), the differentiator and gradient filter use \(y_j\) and \(q_a(t_j^{\rm s,-})\), where \(q_a(t_j^{\rm s,-})=q_{a,j-1}\) for \(j\ge1\). The post-update states generate \(q_{c,j}\), which is held on \([t_j^{\rm s},t_{j+1}^{\rm s})\).
It is initialized with \(\widehat y_{c,0}=y_0\), \(w_{c,0}=g_{l,0}=z_{c,0}=0\), and the first probe direction \(+v_p\). This comparator retains the super-twisting differentiation and optimizer structure of \citet{TorresZuniga2021} and uses a regularized derivative ratio. It additionally receives the measured applied rate \(q_a\); supplying this signal permits the comparison to focus on the regularized-ratio mechanism without rate-estimation error. Define \(r_c:=w_c-J'(\theta)q_a\), which collects its plant-dynamic, sensing, and differentiation errors. Then
\begin{equation}
 w_c=J'(\theta)q_a+r_c,
 \qquad
 g_r=J'(\theta)\frac{q_a^2}{q_a^2+\varepsilon_c^2}
 +r_c\frac{q_a}{q_a^2+\varepsilon_c^2}
 \label{eq:stw_regularization_decomp}
\end{equation}
shows that regularization keeps the ratio bounded but does not create excitation: its gradient multiplier vanishes at \(q_a=0\). 

\subsection{Common scenario distributions and performance metrics}
The notation \(\mathcal U[a,b]\) denotes the uniform distribution on \([a,b]\). The common scenario distribution is
\begin{align*}
 \theta(0)&\sim\mathcal U[0.08,1],\\
 k_1&\sim\mathcal U[0.9,1.1],\qquad
 k_2/0.02\sim\mathcal U[0.9,1.1].
\end{align*}
Training, validation, and held-out cases independently draw the initial feed velocity and kinetic parameters from this distribution. Each case also uses independent sensing phases \(\phi_{\nu,i}\sim\mathcal U[0,2\pi]\), \(i=1,2,3\). Unless stated otherwise, the training, validation, held-out, sampling-period, and initialization studies use \(d_\theta\equiv0\). A nonzero matched disturbance is introduced only in the matched-rate sweep. The held-out evaluation uses 200 paired cases. Each reactor run is initialized at the discrete steady state associated with its sampled initial parameter and kinetics. The smooth sensing disturbance is
\begin{equation}
\begin{split}
 \nu(t)=10^{-3}\{&0.5\sin(0.31t+\phi_{\nu,1})
 +0.3\sin(0.73t+\phi_{\nu,2})\\
 &+0.2\sin(1.37t+\phi_{\nu,3})\},
\end{split}
\label{eq:numerical_noise}
\end{equation}
The horizon is \(T=180\) s and \(\Delta t=5\times10^{-3}\) s. Define the tail-sample index set
\[
 \mathcal J_{\rm tail}:=\{j\in\{0,\ldots,N_s\}:t_j^{\rm s}\in[144,180]\},
 \qquad b_{N,j}:=b_N(t_j^{\rm s}),
\]
and let \([z]_+:=\max\{z,0\}\). The sampled performance metrics are
\begin{align}
 e_J&:=\frac{1}{\#\mathcal J_{\rm tail}}
 \sum_{j\in\mathcal J_{\rm tail}}[b_N^\star-b_{N,j}]_+,
 \label{eq:metric_gap}\\
 e_\theta&:=\max_{j\in\mathcal J_{\rm tail}}|\theta_j-\theta^\star|,
 \label{eq:metric_theta}\\
 q_{\max}&:=\max_{0\le j<N_s}|q_{a,j}|,
 \qquad
 q_{\rm rms}:=\left(\frac{1}{N_s}\sum_{j=0}^{N_s-1}q_{a,j}^2\right)^{1/2}.
 \label{eq:rate_metrics}
\end{align}
The positive part in \eqref{eq:metric_gap} avoids rewarding temporary overshoots above the steady-state optimum. Success requires \(e_J\le0.005\) and \(e_\theta\le0.08\). The two thresholds are about \(0.54\%\) of the nominal optimum and \(7\%\) of the parameter interval, respectively, and are fixed before testing. A compact sensitivity check uses \(e_J^{\rm th}\in\{0.0025,0.005,0.01\}\) and \(e_\theta^{\rm th}\in\{0.04,0.08,0.12\}\). The empirical 90th percentile is denoted by \(Q_{0.9}(\cdot)\).

\subsection{Tuning protocol and selected gains}
For compactness, define the tuning vectors
\[
 p_I:=(\rho,\tau_{\min},h_g,\lambda_1,\lambda_2,T_A,k_I),\quad
 p_P:=(\omega,a_p,k_{\rm esc},r_h,r_l,\phi),
\]
\[
 p_S:=(\lambda_c,\alpha_c,\varepsilon_c,\lambda_{1c},\lambda_{2c},f_g,h_c,v_p,T_p).
\]
Table~\ref{tab:tuning_ranges} states the global search boxes used before testing.

\begin{table*}[t]
\centering\scriptsize
\caption{Global tuning search spaces.}
\label{tab:tuning_ranges}
\begin{tabular}{@{}p{0.16\textwidth}p{0.78\textwidth}@{}}
\toprule
Method & Global search box\\
\midrule
Causal ISM--STW & \(\rho\in[0.004,0.015]\); \(\tau_{\min}\in[0.7,4.5]\); \(h_g\in[0.002,0.08]\); \(\lambda_1\in[0.10,0.55]\); \(\lambda_2\in[0.018,0.18]\); \(T_A\in[0.4,5.0]\); \(k_I\in[0.012,0.085]\).\\
P-ESC-T & \(\omega\in[0.10,1.8]\); \(a_p\in[0.003,0.07]\); \(k_{\rm esc}\in[0.01,3.0]\); \(r_h\in[0.015,0.7]\); \(r_l\in[0.03,1.5]\); \(\phi\in[-\pi,\pi]\).\\
Rate-informed STW & \(\lambda_c\in[0.002,0.06]\); \(\alpha_c\in[5\!\times\!10^{-7},1.5\!\times\!10^{-3}]\); \(\varepsilon_c\in[5\!\times\!10^{-4},0.04]\); \(\lambda_{1c}\in[0.08,0.65]\); \(\lambda_{2c}\in[0.012,0.30]\); \(f_g\in[0.03,4.0]\); \(h_c\in[0.001,0.15]\); \(v_p\in[0.004,0.030]\); \(T_p\in[6,55]\).\\
\bottomrule
\end{tabular}
\end{table*}

In the global search, all coordinates of \(p_I\) are sampled uniformly except for the log-uniform deadband \(h_g\). The five positive coordinates of \(p_P\) are log-uniform and \(\phi\) is uniform. For \(p_S\), \(\lambda_c,\alpha_c,\varepsilon_c,f_g,h_c\) are log-uniform, while the remaining coordinates are uniform. Each tuned method receives 160 global and 160 local evaluations. For positive coordinates, local candidates are generated by independent multiplicative perturbations about the best feasible global candidate and clipped to pre-specified, modestly expanded admissible domains. The Causal ISM--STW local deadband domain extends to \(h_g=0.001\). The P-ESC-T phase receives a pre-specified additive perturbation and is wrapped to \([-\pi,\pi]\).

Tuning uses 16 independent training cases and 100 independent validation cases from the common scenario distribution above. Global candidates are first scored on training. For each tuned method, the best feasible global candidate on the training set seeds the local search; the local candidates are then scored on the same training set. The 24 best feasible candidates from the combined global and local pools are subsequently scored on validation. At both stages, feasibility requires \(Q_{0.9}(\mathcal A_\theta)\le0.015\), \(Q_{0.9}(p_{\partial\Theta})=0\), and \(Q_{0.9}(p_{\rm cap})=0\). Among feasible candidates, selection lexicographically minimizes
\[
 \bigl(-\text{success rate},\ \operatorname{median}(e_J),\
 Q_{0.9}(e_J),\ \operatorname{median}(\mathcal A_\theta)\bigr).
\]
Table~\ref{tab:tuning} reports the selected values.

\begin{table*}[t]
\centering\scriptsize
\caption{Controller parameters selected on validation and frozen for all held-out and sensitivity tests.}
\label{tab:tuning}
\begin{tabular}{@{}p{0.17\textwidth}p{0.76\textwidth}@{}}
\toprule
Method & Selected values\\
\midrule
Causal ISM--STW & $\rho=0.0107882$, $\tau_{\min}=0.814674$, $h_g=0.0019813$, $(\lambda_1,\lambda_2)=(0.36755,0.11173)$, $T_A=0.940735$, $k_I=0.0207689$\\
P-ESC-T & $(\omega,a_p,k_{\rm esc})=(0.582654,0.0316391,2.27289)$, $(r_h,r_l)=(0.0359860,0.0701390)$, $\phi=-0.482157$\\
Rate-informed STW & $(\lambda_c,\alpha_c,\varepsilon_c)=(0.0186518,5.09507\!\times\!10^{-7},0.00649018)$, $(\lambda_{1c},\lambda_{2c})=(0.236359,0.0829409)$, $f_g=0.824085$, $h_c=0.0266345$, $v_p=0.00717757$, $T_p=20.2447$\\
\bottomrule
\end{tabular}
\end{table*}

\subsection{Held-out results}
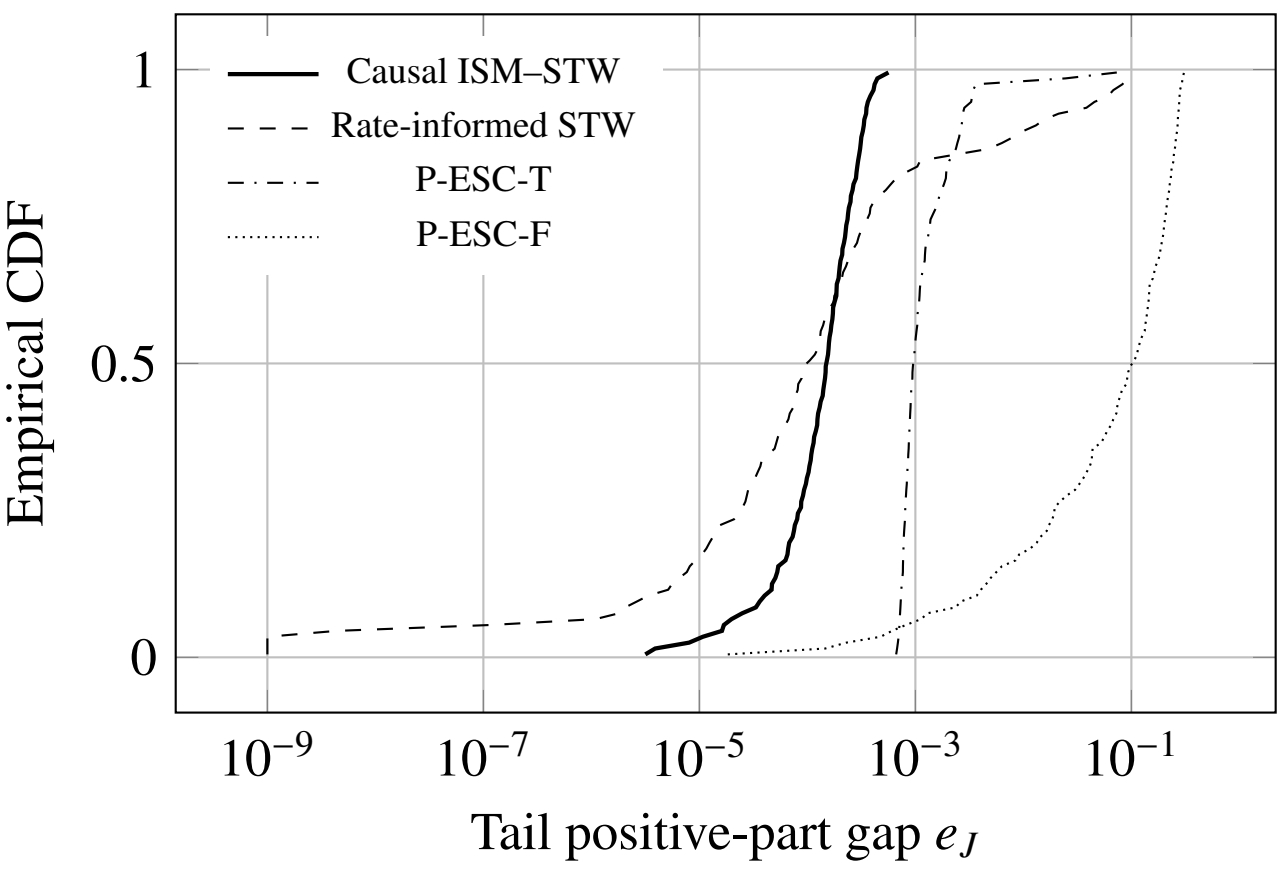
\begin{figure}[t]
\centering
\begin{tikzpicture}
\begin{axis}[width=\columnwidth,height=0.70\columnwidth,xmode=log,grid=major,
 xlabel={Tail positive-part gap $e_J$},ylabel={Empirical CDF},
 legend style={font=\scriptsize,at={(0.03,0.97)},anchor=north west,draw=none}]
 \addplot[black,thick] table[x=proposed,y=p]{\heldoutcdfdata};\addlegendentry{Causal ISM--STW}
 \addplot[black,dashed] table[x=stw_gradient,y=p]{\heldoutcdfdata};\addlegendentry{Rate-informed STW}
 \addplot[black,dashdotted] table[x=pesc_tuned,y=p]{\heldoutcdfdata};\addlegendentry{P-ESC-T}
 \addplot[black,densely dotted] table[x=pesc_fixed,y=p]{\heldoutcdfdata};\addlegendentry{P-ESC-F}
\end{axis}
\end{tikzpicture}
\caption{Empirical CDFs of \(e_J\) in the held-out cases. Zero gaps are displayed at \(10^{-9}\) solely for the logarithmic axis. The proposed method gives the smallest upper tail; the rate-informed STW comparator is more accurate near the median but has a heavier failure tail.}
\label{fig:heldout_cdf}
\end{figure}

Table~\ref{tab:heldout} summarizes the paired results, and Figure~\ref{fig:heldout_cdf} shows their empirical gap distributions. The proposed controller has the highest success rate and the smallest 90th-percentile gap. The rate-informed STW comparator has a smaller overall median gap and lower activity, but develops a substantially heavier upper tail. P-ESC-T has about 6.4 times the proposed median gap and a similar median RMS rate. For the causal ISM--STW, rate-informed STW, and P-ESC-F methods, the common boundary and rate-cap operators remain inactive in all held-out cases; P-ESC-T activates them only in a small number of cases, with zero activation at the 90th percentile.

\begin{table*}[t]
\centering\scriptsize
\caption{Two hundred paired held-out Monte Carlo cases.}
\label{tab:heldout}
\begin{tabular}{lccccccc}
\toprule
Method & Success & Median $e_J$ & $Q_{0.9}(e_J)$ & Median $e_\theta$ & Median $\mathcal A_\theta$ & Median $q_{\max}$ & Median $q_{\rm rms}$\\
\midrule
Causal ISM--STW & 200/200 & $1.49\!\times\!10^{-4}$ & $3.28\!\times\!10^{-4}$ & 0.0167 & 0.0108 & 0.0108 & 0.0108\\
Rate-informed STW & 174/200 & $9.88\!\times\!10^{-5}$ & $1.11\!\times\!10^{-2}$ & 0.0069 & 0.0022 & 0.0108 & 0.0044\\
P-ESC-T & 196/200 & $9.57\!\times\!10^{-4}$ & $2.61\!\times\!10^{-3}$ & 0.0458 & 0.0101 & 0.0220 & 0.0113\\
P-ESC-F & 28/200 & $1.02\!\times\!10^{-1}$ & $2.66\!\times\!10^{-1}$ & 0.2594 & 0.0003 & 0.0004 & 0.0003\\
\bottomrule
\end{tabular}
\end{table*}

\subsection{Matched-rate sweep and nonvanishing-motion diagnostics}
With all gains frozen, 100 new paired cases are evaluated at each disturbance-to-search ratio. For independent phases \(\varphi_i\sim\mathcal U[0,2\pi]\), \(i=1,2\), define
\begin{align}
 d_\theta^{(\kappa)}(t)&=\kappa\rho\,\psi(t;\varphi_1,\varphi_2),\nonumber\\
 \psi(t;\varphi_1,\varphi_2)&:=0.65\sin(0.17t+\varphi_1)
 +0.35\sin(0.53t+\varphi_2).
 \label{eq:rate_dist_sweep}
\end{align}
Here \(\rho=0.0107882\) is the selected Causal ISM--STW search speed. The resulting disturbance signal is applied unchanged to every method through the common parameter-rate channel. The sweep uses \(\kappa\in\{0,0.5,0.9,1.1,1.3\}\), with the same scenarios and phases for every method and level. The selected gain satisfies \(k_I>1.3\rho\); the causal realization is assessed through its measured rate mismatch. For the No-ISM ablation, the complete pre-safeguard rate is \(\widetilde q_{a,j}=v_j^+ + d_{\theta,j}^{(\kappa)}\), after which the common rate cap and interval safeguard are applied.

Using the sample-count notation defined above, let \(\mathcal J_{\rm eval}\subset\{0,\ldots,N_s-1\}\) contain the post-activation update intervals for which the common safeguards are inactive, and fix \(\delta_s=0.1\). The set \(\mathcal J_{\rm eval}\) is nonempty in every reported simulation case, so the following metrics are well defined. The prescribed search velocity held on interval \(j\) is \(v_j^+\). Define
\begin{align}
 p_{\rm stall}&:=\frac{\#\{j\in\mathcal J_{\rm eval}:|q_{a,j}|\le\delta_s\rho\}}
 {\#\mathcal J_{\rm eval}},
 \label{eq:pstall}\\
 p_{\rm opp}&:=\frac{\#\{j\in\mathcal J_{\rm eval}:v_j^+q_{a,j}<0\}}
 {\#\mathcal J_{\rm eval}},
 \label{eq:popp}\\
 m_v&:=\frac{1}{\#\mathcal J_{\rm eval}}
 \sum_{j\in\mathcal J_{\rm eval}}\frac{|q_{a,j}-v_j^+|}{\rho}.
 \label{eq:rate_mismatch_metric}
\end{align}
The quantity \(m_v\) detects magnitude distortion even when complete stall or reversal does not occur. On the nominal equilibrium map, the smaller gradient magnitude at the parameter-error success boundary \(|\theta-\theta^\star|=0.08\) is 0.2795. The threshold 0.05, fixed before the sweep, is therefore less than one fifth of that value and excludes only the near-optimum small-gradient region. The regularization multiplier of the rate-informed STW comparator is evaluated after its probe on samples with inactive safeguards and \(|J'(\theta_j)|\ge0.05\); it falls below one half on 29--78\% of those samples across the sweep.

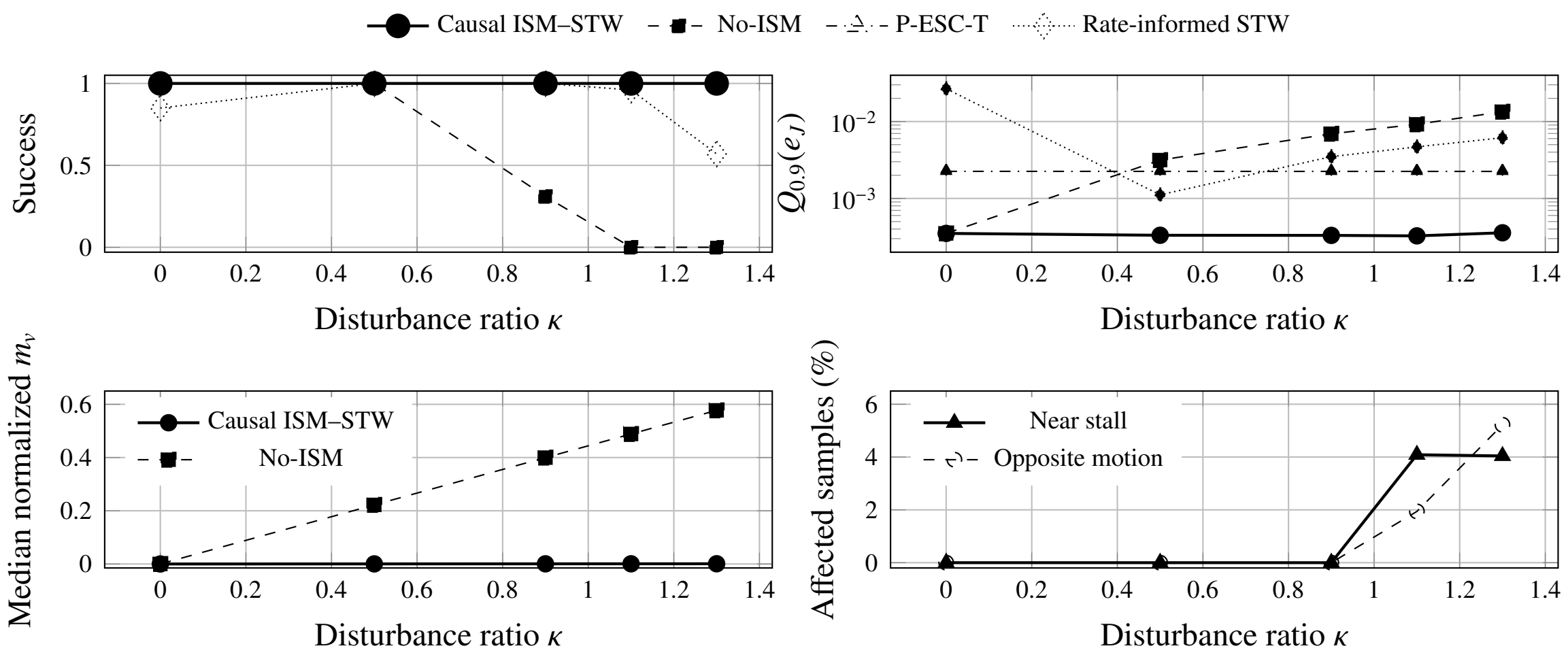
\begin{figure*}[t]
\centering
\begin{tikzpicture}
\begin{groupplot}[
 group style={group name=rateplots,group size=2 by 2,horizontal sep=1.15cm,vertical sep=1.35cm},
 width=0.44\textwidth,height=0.18\textwidth,
 grid=major,tick label style={font=\scriptsize},label style={font=\small},
 legend style={font=\scriptsize,draw=none}]
\nextgroupplot[xlabel={Disturbance ratio $\kappa$},ylabel={Success},
 ymin=-0.03,ymax=1.05,
 legend to name=ratelegend,legend columns=4,
 legend style={font=\scriptsize,draw=none,
 /tikz/every even column/.append style={column sep=0.45em}}]
 \addplot[black,thick,mark=*,mark size=3.0pt] table[x=kappa,y=prop_success]{\ratesweepdata};\addlegendentry{Causal ISM--STW}
 \addplot[black,dashed,mark=square*,mark size=1.8pt] table[x=kappa,y=noism_success]{\ratesweepdata};\addlegendentry{No-ISM}
 \addplot[black,dashdotted,mark=triangle,mark size=3.0pt] table[x=kappa,y=pesc_success]{\ratesweepdata};\addlegendentry{P-ESC-T}
 \addplot[black,densely dotted,mark=diamond,mark size=3.5pt] table[x=kappa,y=stw_success]{\ratesweepdata};\addlegendentry{Rate-informed STW}
\nextgroupplot[xlabel={Disturbance ratio $\kappa$},ylabel={$Q_{0.9}(e_J)$},
 ymode=log,ymin=2e-4,ymax=4e-2]
 \addplot[black,thick,mark=*] table[x=kappa,y=prop_q90]{\ratesweepdata};
 \addplot[black,dashed,mark=square*] table[x=kappa,y=noism_q90]{\ratesweepdata};
 \addplot[black,dashdotted,mark=triangle*] table[x=kappa,y=pesc_q90]{\ratesweepdata};
 \addplot[black,densely dotted,mark=diamond*] table[x=kappa,y=stw_q90]{\ratesweepdata};
\nextgroupplot[xlabel={Disturbance ratio $\kappa$},ylabel={Median normalized $m_v$},
 ymin=-0.015,ymax=0.65,legend pos=north west]
 \addplot[black,thick,mark=*] table[x=kappa,y=prop_mismatch]{\ratesweepdata};\addlegendentry{Causal ISM--STW}
 \addplot[black,dashed,mark=square*] table[x=kappa,y=noism_mismatch]{\ratesweepdata};\addlegendentry{No-ISM}
\nextgroupplot[xlabel={Disturbance ratio $\kappa$},ylabel={Affected samples (\%)},
 ymin=-0.2,ymax=6.5,legend pos=north west]
 \addplot[black,thick,mark=triangle*] table[x=kappa,y=noism_stall_pct]{\ratesweepdata};\addlegendentry{Near stall}
 \addplot[black,dashed,mark=o] table[x=kappa,y=noism_opp_pct]{\ratesweepdata};\addlegendentry{Opposite motion}
\end{groupplot}
\path (rateplots c1r1.north west)--(rateplots c2r1.north east)
 node[midway,above=3pt] {\pgfplotslegendfromname{ratelegend}};
\end{tikzpicture}
\caption{Matched-rate performance and applied-motion diagnostics with 100 paired cases per level. Top left: success rate. Causal ISM--STW and P-ESC-T coincide at unit success throughout; No-ISM also coincides with them at $\kappa=0$ and $0.5$, and Rate-informed STW at $\kappa=0.5$ and $0.9$. Top right: upper-tail objective gap. Bottom left: median normalized applied-rate mismatch. Bottom right: fractions of safeguard-inactive No-ISM post-activation samples in near-stall and opposite-motion regimes.}
\label{fig:rate_sweep_diagnostics}
\end{figure*}

Figure~\ref{fig:rate_sweep_diagnostics} reports the sweep. The causal ISM--STW succeeds in all 100 cases at every \(\kappa\), its 90th-percentile gap remains below \(3.6\times10^{-4}\), and its measured \(p_{\rm stall}\) and \(p_{\rm opp}\) are zero, while the median \(m_v\) remains below \(8.4\times10^{-4}\). For the No-ISM ablation, the median normalized mismatch is 0.400 at \(\kappa=0.9\), where success decreases to 31\% even though complete stall and reversal have not yet appeared. For \(\kappa>1\), near-stall and opposite-motion intervals also occur and the No-ISM success rate is zero. P-ESC-T retains 100\% success in this sweep but has an upper-tail gap near \(2.25\times10^{-3}\). The rate-informed STW comparator exhibits conditional attenuation over 29--78\% of its nontrivial-gradient samples across the sweep, and its success rate decreases to 57\% at \(\kappa=1.3\). The sweep therefore associates No-ISM degradation first with increasing applied-rate mismatch and, beyond unit disturbance ratio, with near-stall and opposite motion.

\subsection{Sampled realization, sensitivity, and analytical diagnostics}
\label{sec:diagnostics}
The main simulations use \eqref{eq:causal_proximal_ism}. A paired sensitivity study over \(\Delta t\in\{0.01,0.005,0.0025\}\,\mathrm{s}\) compares it with \eqref{eq:explicit_sampled_ism} using frozen gains. Both realizations succeed in 48--50 of the 50 cases at each sampling period. Their median objective gaps remain of the same order, while the proximal realization consistently has lower activity (median \(\mathcal A_\theta=0.0108\) versus \(0.0195\)). Because relay decisions change with the sampling grid, this is a sampling-period sensitivity study rather than a numerical-order test.

The proposed method remains successful in all 200 held-out cases for all nine tested threshold pairs; no comparator achieves 200/200 throughout this grid. The initialization study compares the proposed method with the two tuned comparators, P-ESC-T and Rate-informed STW, under 5\% and 10\% smooth perturbations of the equilibrium profiles. For each paired case, a phase \(\phi_{\rm init}\sim\mathcal U[0,2\pi]\) is shared across methods. With \(z_i=(i-\tfrac12)/N\) and \(a_{\rm init}\in\{0,0.05,0.10\}\), the steady-state profiles are multiplied by \(1+a_{\rm init}\sin(2\pi z_i+\phi_{\rm init})\) for \(a_i\) and \(1-a_{\rm init}\cos(2\pi z_i+\phi_{\rm init})\) for \(b_i\), with negative values clipped at zero. The proposed method succeeds in all 50 paired cases at each level. At 0\%, 5\%, and 10\% perturbation, P-ESC-T succeeds in 50/50, 43/50, and 37/50 cases, while Rate-informed STW succeeds in 44/50 cases at each level.

Forty independent diagnostic runs with \(d_\theta\equiv0\) use the same scenario and sensing-phase distributions. The derivative error compares the post-update differentiator estimate with the same-interval forward difference after the dwell-based recovery window. The endpoint margin is evaluated at a curvature test level 25\% above the largest observed finite-difference curvature with a trace-normalized differentiator matrix, \(\operatorname{trace}P_{D,\mathrm r}=1\). The largest recovered derivative error is \(1.12\times10^{-3}\), and, for the proposed controller, the common safeguards remain inactive in all 200 held-out runs. These reactor quantities are empirical mechanism diagnostics.

\begin{table}[t]
\centering
\caption{Empirical mechanism diagnostics for the selected Causal ISM--STW gains.}
\label{tab:reactor_diagnostics}
\scriptsize
\begin{tabular}{p{0.64\columnwidth}c}
\toprule
Quantity & Value / status\\
\midrule
Rate-authority ratio at \(\kappa_{\max}=1.3\), \(k_I/(\kappa_{\max}\rho)\) & 1.48\\
Endpoint-LMI margin & \(4.34\times10^{-2}\)\\
Maximum recovered derivative error & \(1.12\times10^{-3}\)\\
Maximum proposed-method \((p_{\partial\Theta},p_{\rm cap})\) & \((0,0)\)\\
\bottomrule
\end{tabular}
\end{table}

A scalar special case verifies simultaneous feasibility of the theorem conditions. Consider \(\dot x=-20(x-\theta)\), \(h(x)=\tfrac12(x-1)^2\), and \(\Theta=[0.5,1.5]\), for which \(\ell(\theta)=\theta\), \(J(\theta)=\tfrac12(\theta-1)^2\), \(\alpha_J(r)=r\), and \(L_J=1\). Use \(\theta(0)=x(0)=1.1\), \((\rho,\rho_0)=(0.01,0.02)\), \((r_x,r_\eta)=(0.02,0.01)\), \((\tau_{\min},h_g)=(0.5,0.005)\), \((k_I,\overline d_\theta)=(0.02,0.005)\), \((\varepsilon_s,C_V)=(1,3\times10^{-4})\), and \((\lambda_1,\lambda_2)=(10,8)\), with \(\nu\equiv0\), \(v_0=-\rho\), and the standard differentiator and integral-sliding initializations. Taking the scalar plant Lyapunov matrix \(P_{\rm s}=1/40\), \(\tau_0=0.3821\), and \(T_A=T_D=3.8476\), note that the standard integral initialization gives \(\lvert\dot\theta\rvert=\rho\) from \(t=0\). The scalar differentiator certificate therefore uses this sharpened rate bound when evaluating \(Y_2\). Since \(\alpha_J(r)=r\), this example has \(R_\rho=\varepsilon_{\rm tot}\). A feasible differentiator certificate is
\[
 P_{D,\mathrm s}=
 \begin{bmatrix}
 0.838107&-0.201225\\
 -0.201225&0.161893
 \end{bmatrix},
 \qquad
 Q_{D,\mathrm s}=0.09660I.
\]

\begin{table}[ht]
\centering
\caption{Complete analytical certificate for the scalar special case.}
\label{tab:scalar_certificate}
\scriptsize
\begin{tabular}{lc}
\toprule
Condition or quantity & Value\\
\midrule
$q_s$ & $9.08\times10^{-5}$\\
$\tau_r$ / $\tau_{\min}$ & $0.4861$ / $0.5000$\\
Local-consistency left / right side & $5.00\times10^{-8}$ / $2.50\times10^{-6}$\\
$C_\Delta$ & 1.1915\\
$\varepsilon_{\rm tot}$ / $R_\rho$ & 0.01691 / 0.01691\\
Parameter-error / gradient bound & 0.02191 / 0.02191\\
$T_D$ / $T_A$ & $3.8476$ / $3.8476$\\
$\operatorname{dist}(\mathcal I_\Theta(\theta(0)),\partial\Theta)$ & 0.3396\\
$Y_2$ / endpoint-LMI margin & 4.424 / 0.0966\\
$k_I-\overline d_\theta$ & 0.015\\
\bottomrule
\end{tabular}
\end{table}

Each numerical test has a distinct role. The held-out study evaluates optimization accuracy and activity under fixed gains, the matched-rate sweep isolates the role of nonvanishing applied motion, and the sensitivity tests assess implementation robustness. The reactor diagnostics check the mechanisms used by the analysis.

\section{Conclusion}
An output-feedback ISM--STW extremum seeker was developed for stabilized nonlinear systems with unknown equilibrium maps. Within the local state--parameter domain, integral sliding enforces the prescribed nonzero parameter rate after capture. The ramp-corrected analysis yields an order-\(\rho^2\) plant remainder, and local regularity bounds establish finite-time exact STW differentiation. The resulting practical parameter-error and gradient bounds separate dynamic lag, sensing, deadband, and dwell effects.

On the 80-state reactor benchmark, the causal realization succeeded in all 200 held-out cases and attained the smallest upper-tail gap. The matched-rate sweep associated No-ISM degradation with increasing rate mismatch, followed by near-stall motion and direction reversal.

The proposed method succeeded for every tested threshold pair and initialization level. The scalar case verified simultaneous feasibility of the continuous-time conditions. A sampled-data analysis is needed to turn the observed sampling-period robustness into a convergence guarantee.

\section*{Declaration of competing interest}
The author declares no known competing financial interests or personal relationships that could have appeared to influence the work reported in this paper.

\end{document}